\documentclass[12pt]{amsart}

\copyrightinfo{2000}{American Mathematical Society}

\newcommand{\cA}{{\mathcal A}}
\newcommand{\cB}{{\mathcal B}}
\newcommand{\cD}{{\mathcal D}}
\newcommand{\cE}{{\mathcal E}}
\newcommand{\cL}{{\mathcal L}}
\newcommand{\cS}{{\mathcal S}}

\newcommand{\bN}{{\mathbb N}}
\newcommand{\bZ}{{\mathbb Z}}
\newcommand{\bQ}{{\mathbb Q}}
\newcommand{\bR}{{\mathbb R}}
\newcommand{\bC}{{\mathbb C}}

\numberwithin{equation}{section}

\newtheorem{Theorem}{Theorem}[section]
\newtheorem{Lemma}{Lemma}[section]
\newtheorem{Corollary}{Corollary}[section]
\newtheorem{Definition}{Definition}[section]
\newtheorem{Remark}{Remark}[section]
\newtheorem{Example}{Example}[section]
\newtheorem{Proposition}{Proposition}[section]

\author{S.~Albeverio}
\address{Institut f\"{u}r Angewandte Mathematik, Universit\"{a}t Bonn,
Wegelerstra\ss e 6, D-53115 Bonn, Germany; \ SFB 611, Bonn, \
BiBoS, Bielefeld -- Bonn}
\email{albeverio@uni-bonn.de}

\author{A.~Yu.~Khrennikov}
\address{International Center for Mathematical Modelling in Physics
and Cognitive Sciences MSI, V\"axj\"o University, \ SE-351 95,
V\"axj\"o, \ Sweden.}
\email{andrei.khrennikov@msi.vxu.se}

\author{V.~M.~Shelkovich}
\address{Department of Mathematics, St.-Petersburg State Architecture
and Civil Engineering University, \ 2 Krasnoarmeiskaya 4, 190005,
St. Petersburg, \ Russia.}
\email{shelkv@vs1567.spb.edu}

\title[$p$-Adic analysis in the Lizorkin type spaces]
{$p$-Adic analysis in the Lizorkin type spaces: \\
fractional operators,
pseudo-differential equations and Tauberian theorems}

\thanks{This paper was supported in part by the grant
of The Swedish Royal Academy of Sciences on collaboration with scientists of
former Soviet Union and the EU-Network ``Quantum Probability and
Applications''.
The first and the third authors (S.~A. and V.~S.) were also supported
in part by DFG Project 436 RUS 113/809/0-1.}



\date{ }

\begin{document}

\begin{abstract}
In this paper the $p$-adic Lizorkin spaces of test functions
and distributions are introduced, and multidimensional Vladimirov's
and Taibleson's fractional operators are studied on these spaces.
Since the $p$-adic Lizorkin spaces are invariant under the
Vladimirov and Taibleson operators, they can play a key role in
considerations related to fractional operator problems.
A class of $p$-adic pseudo-differential operators in the Lizorkin
spaces is also introduced and solutions of pseudo-differential
equations are constructed.
$p$-Adic multidimensional Tauberian theorems connected with fractional
operators and pseudo-differential operators for the Lizorkin distributions
are also proved.
\end{abstract}

\maketitle

\section{Introduction}
\label{s1}

\subsection{$p$-Adic mathematical physics.}\label{s1.1}
It is well known that apart from the {\it ``usual'' mathematical physics\/}
(``$\bC$--case'', where all functions and distributions are complex or
real valued defined on spaces with real or complex coordinates)
there is a {\it $p$-adic mathematical physics\/} where all functions
and distributions are defined on the field $\bQ_p$ of $p$-adic numbers
(definition of the field $\bQ_p$ see below in Sec.~\ref{s2}).

There are a lot of papers where different applications of
$p$-adic analysis to physical problems (in the strings theory, in quantum mechanics),
stochastics, in the theory of dynamical systems, cognitive sciences
and psychology are studied~\cite{Al-Gor-Kh}--\cite{Al-Kh1},
~\cite{Al-Kh-Ti}--\cite{Bik-V},~\cite{Ev},~\cite{Fr-W1},~\cite{Kh1}--~\cite{Kh5},
~\cite{Koch3},~\cite{Par},~\cite{Vl-V-Z}--~\cite{V2} (see also the
references therein).
Note that the theory of $p$-adic distributions (generalized functions)
plays an important role in solving mathematical problems of
$p$-adic analysis and applications. Fundamental results about the
$p$-adic theory of distributions can be found in~\cite{Br},~\cite{G-Gr-P},
~\cite{Kh1},~\cite{Taib3},~\cite{Vl-V-Z}).
Note that to deal with {\it nonlinear singular problems\/} of $p$-adic
mathematical physics, in~\cite{Al-Kh-Sh3}--~\cite{Al-Kh-Sh5} algebraic
nonlinear theories of distributions were constructed.

Since there exists a $p$-adic analysis connected with the mapping
$\bQ_p$ into $\bQ_p$ and an analysis connected with the mapping $\bQ_p$
into the field of complex numbers $\bC$, there exist two types of
$p$-adic physics models.
It is known that for the $p$-adic analysis related to the mapping
$\bQ_p \to \bC$, the operation of partial differentiation is
{\it not defined\/}, and the Vladimirov fractional operator
$D^{\alpha}=f_{-\alpha}*$ plays a corresponding role~\cite[IX]{Vl-V-Z}, where
$f_{\alpha}$ is the $p$-adic {\it Riesz kernel\/} (\ref{56}), $*$ is a
convolution. Moreover, large quantity of $p$-adic models use the Vladimirov
fractional operator and the theory of $p$-adic distributions
~\cite{Al-K},~\cite{Al-Kh1},~\cite{Av-Bik-Koz-O},~\cite{Bik-V},
~\cite{Fr-W1},~\cite{Kh1}--~\cite{Kh4},~\cite{Koch3},~\cite{Par},~\cite{Vl-V-Z};
further generalizations can be found in~\cite{Koz1},~\cite{Koz2}.
However, in general, $D^{\alpha}\varphi \not\in {\cD}(\bQ_p)$ for
$\varphi \in {\cD}(\bQ_p)$, and consequently, the operation $D^{\alpha}f$
is well defined only for some distributions $f\in {\cD}'(\bQ_p)$.
For example, in general, $D^{-1}$ is not defined in the space of
test functions ${\cD}(\bQ_p)$~\cite[IX.2]{Vl-V-Z}.

We recall that similar problems arise for the ``$\bC$-case'' of
fractional operators~\cite{Rub}, \cite{Sam3}, \cite{Sam-Kil-Mar}. Namely,
in general, the Schwartzian test function space ${\cS}(\bR^n)$
{\it is not invariant\/} under fractional operators.
A solution of this problem (in the ``$\bC$-case'') was suggested by
P.~I.~Lizorkin in the excellent papers~\cite{Liz1}--~\cite{Liz3}
(see also~\cite{Sam1},~\cite{Sam2}).
Namely, in~\cite{Liz1}--~\cite{Liz3} a new type spaces
{\it invariant\/} under fractional operators were introduced.

We recall the definition of one type of the Lizorkin space
(for details, see~\cite{Liz3}, \cite{Sam3}, \cite{Sam-Kil-Mar}).
Denote by $\bN$, $\bR$, $\bC$ the sets of
positive integers, real numbers and complex
numbers, respectively, and set ${\bN}_0={0}\cup{\bN}$. For
$\alpha=(\alpha_1,\dots,\alpha_n)\in {\bN}_0^n$ and
$x=(x_1,\dots,x_n)\in \bR^n$ we assume
$|\alpha|=\sum_{k=1}^n\alpha_k$ and
$x^{\alpha}=x_1^{\alpha_1}\cdots x_n^{\alpha_n}$. We shall
denote partial derivatives of the order $|\alpha|$ by
$\partial_x^{\alpha}=\frac{\partial^{|\alpha|}}
{\partial{x_1}^{\alpha_1}\cdots\partial{x_n}^{\alpha_n}}$.
Now let us consider the following subspace of test functions
\begin{equation}
\label{5}
\Psi(\bR^n)=\{\psi(\xi): \psi\in \cS(\bR^n):
(\partial_{\xi}^{j}\psi)(0)=0, \, |j|=1,2,\dots\},
\end{equation}
The space of functions
\begin{equation}
\label{6}
\Phi(\bR^n)=\{\phi: \phi=F[\psi], \, \psi\in \Psi(\bR^n)\}
\subset \cS(\bR^n),
\end{equation}
is called the {\it Lizorkin space\/}, where $F$ is the Fourier transform.
This space admits a simple characterization: $\phi\in \Phi(\bR^n)$
if and only if $\phi\in \cS(\bR^n)$ and
\begin{equation}
\label{6.1}
\int_{\bR^n}x^j\phi(x)\,d^nx=0, \quad |j|=0,1,2,\dots.
\end{equation}
Thus $\Phi(\bR^n)$ is the subspace of Schwartzian test functions,
for which all the moments are equal to zero.

It is well known that $\Phi(\bR^n)$ is invariant under the Riesz
fractional operator $D^{\alpha}$, $\alpha \in \bC$, given by the formula
\begin{equation}
\label{7}
\big(D^{\alpha}\phi\big)(x)
\stackrel{def}{=}(-\Delta)^{\alpha/2}\phi(x)
=\kappa_{-\alpha}(x)*\phi(x), \quad \phi \in \Phi(\bR^n),
\end{equation}
where the {\it Riesz kernel\/} is defined as
$\kappa_{\alpha}(x)
=\frac{\Gamma(\frac{n-\alpha}{2})}
{2^{\alpha}\pi^{\frac{n}{2}}\Gamma(\frac{\alpha}{2})}|x|^{\alpha-n}$,
where $|x|=\sqrt{x_1^2+\cdots+x_n^2}$ and $|x|^{\alpha}$ is a homogeneous
distribution of degree~$\alpha$, \ $\Delta$ is the Laplacian.

Note that fractional operators in the ``$\bC$-case'',
as well as in the $p$-adic case have many applications and are intensively
used in mathematical physics~\cite{Eid-Koch}, \cite{Sam3},~\cite{Sam-Kil-Mar}.
These two last fundamental books have the exhaustive references.

We recall also that in the ``$\bC$-case'' the {\it Tauberian theorems\/}
have numerous applications, in particular, in mathematical physics.
Tauberian theorems are usually assumed to connect the asymptotic
behavior of a function (distribution) at zero with asymptotic
behavior of its Fourier, Laplace or other integral transform at infinity.
The inverse theorems are usually called ``Abelian'' \cite{D-Zav1},
 \cite{Kor},~\cite{Vl-D-Zav} (see also the references cited therein).
Multidimensional Tauberian theorems for distributions
are treated in the fundamental book~\cite{Vl-D-Zav}, some of them
are connected with the fractional operator. In~\cite{Vl-D-Zav},
as a rule, theorems of this
type were proved for distributions whose supports belong to
a cone in $\bR^n$ (semiaxis for $n=1$). This is related to the
fact that such distributions form a convolution algebra.
In this case the kernel of the fractional operator is a distribution
whose support belongs to the cone in $\bR^n$ or a semiaxis for $n=1$
~\cite[\S2.8.]{Vl-D-Zav}.

\subsection{Contents of the paper.}\label{s1.2}
In this paper the $p$-adic Lizorkin type spaces and multidimensional
fractional operators and pseudo-differential operators
on these spaces are constructed. Since the Lizorkin
spaces are invariant under fractional operators, they are ``natural''
definition domains of them, and can play
a key role in models related to the fractional operators problems.

In this paper we also prove $p$-adic analogs of Tauberian theorems
for the Lizorkin distributions. Tauberian theorems of this type are
connected with the fractional operators.
Taking into account the fact that kernels of the fractional
operators are defined on the whole space $\bQ_p^n$ (by virtue of the
$p$-adic field nature), Tauberian theorems proved in this paper
are not direct analogs of Tauberian theorems from~\cite{Vl-D-Zav}.
Some $p$-adic Tauberian theorems for distributions in ${\cD}'(\bQ_p^n)$
were first proved in~\cite{Kh-Sh1},~\cite{Kh-Sh2}. Since the space of
distributions ${\cD}'(\bQ_p^n)$ is not invariant under Vladimirov's
operator, mentioned Tauberian theorems in~\cite{Kh-Sh1},~\cite{Kh-Sh2}
have been proved only under reasonable restrictions.
In this respect the present paper gives a more natural framework
for such results.

In Sec.~\ref{s2}, we recall some facts from the $p$-adic
theory of distributions.

In Subsec.~\ref{s3.1} we introduce the $p$-adic Lizorkin spaces of test
functions $\Phi_{\times}(\bQ_p^n)$ and distributions
$\Phi_{\times}'(\bQ_p^n)$ of the first kind, and in Subsec.~\ref{s3.2}
the $p$-adic Lizorkin spaces of test functions $\Phi(\bQ_p^n)$ and
distributions $\Phi'(\bQ_p^n)$ of the second kind. It is easy to see
that the $p$-adic Lizorkin space $\Phi(\bQ_p^n)$ is an analog of
the Lizorkin space $\Phi(\bR^n)$ defined by (\ref{6}).
The Lizorkin spaces $\Phi_{\times}(\bQ_p^n)$ and $\Phi(\bQ_p^n)$
admit characterizations (\ref{50}) and (\ref{54}), respectively.
In Subsec.~\ref{s3.3}, by Lemmas~\ref{lem2.2},~\ref{lem2.3}, we prove
that the Lizorkin spaces $\Phi_{\times}(\bQ_p^n)$ and $\Phi(\bQ_p^n)$
are dense in $\cL^{\rho}(\bQ_p^n)$, $1<\rho<\infty$.
In fact, for $n=1$ and $\rho=2$ this statement was proved
in~\cite[IX.4.]{Vl-V-Z}. Note that for $\rho=2$ the statements
of Lemmas~\ref{lem2.2},~\ref{lem2.3} are almost obvious, but
for $\rho\ne 2$, similarly as for the ``$\bC$-case''~\cite{Sam3},
these statements are nontrivial. Our proofs of these lemmas
almost word for word follow the proofs developed for the
``$\bC$-case'' in~\cite{Sam3}.

In Sec.~\ref{s4} two types of the multidimensional fractional
operators are constructed.
In Subsec.~\ref{s4.1}, we recall some facts on the Vladimirov
one- dimensional fractional operator and introduce the Vladimirov
multidimensional operator $D^{\alpha}_{\times}$ as the direct product
of one-dimensional fractional Vladimirov's operators $D^{\alpha_j}_{x_j}$.
Next, we define this operator in the Lizorkin space of distributions
$\Phi_{\times}'(\bQ_p^n)$ for all $\alpha\in \bC^n$.
In Subsec.~\ref{s4.2} we recall some facts on the multidimensional
fractional operator $D^{\alpha}_{x}$ introduced by
Taibleson~\cite[\S2]{Taib1},~\cite[III.4.]{Taib3}
in the space of distributions ${\cD}'(\bQ_p^n)$ for $\alpha\in \bC$,
$\alpha\ne -n$ and define this operator in the Lizorkin space of
distributions $\Phi'(\bQ_p^n)$ for all $\alpha\in \bC$.
The Lizorkin space $\Phi_{\times}(\bQ_p^n)$ is invariant under
the Vladimirov fractional operator (Lemma~\ref{lem4.1}), while
the Lizorkin space $\Phi(\bQ_p^n)$ is invariant under the Taibleson
fractional operator (Lemma~\ref{lem4.1}).
These fractional operators form Abelian groups on
on the corresponding the Lizorkin spaces (see (\ref{63})).

In fact, in order to define the one-dimensional fractional Vladimirov
operators $D^{-1}$, the one-dimensional Lizorkin space of test functions
$\Phi(\bQ_p)$ was introduced in~\cite[IX.2]{Vl-V-Z} (compare with (\ref{54})).
For $n=1$, according to~\cite[IX,(5.7),(5.8)]{Vl-V-Z} and~\cite{Koz0},
the eigenfunctions (\ref{62.1}) of Vladimirov's operator $D^{\alpha}$,
$\alpha>0$ satisfy condition (\ref{54}), and, consequently, belong
to the Lizorkin space $\Phi(\bQ_p)$.
Moreover, our results imply that these functions (\ref{62.1}) are also
eigenfunctions of the operator $D^{\alpha}$ for $\alpha<0$ (see Remark~\ref{rem1}).

In Subsec.~\ref{s4.3}, by analogy with the ``$\bC$-case'' ~\cite{Sam3},
~\cite{Sam-Kil-Mar}, two types of $p$-adic Laplacians are discussed.
Note that such types of $p$-adic Laplacians were introduced in~\cite{Kh0}.

In Sec.~\ref{s5}, a class of pseudo-differential operators $A$
(\ref{64.3}) on the Lizorkin spaces are introduced. The Lizorkin spaces
are {\it invariant\/} under our pseudo-differential operators.
The fractional operator $D^{\alpha}_{x}$, $\alpha\in \bC$ belongs to
this class of pseudo-differential operators.
The family of pseudo-differential operators $A$ with symbols
$\cA(\xi)\ne 0$, $\xi\in \bQ_p^n\setminus \{0\}$ forms an Abelian group.
In this subsection solutions of pseudo-differential equations
$Af=g$, $g\in \Phi'(\bQ_p^n)$ are also constructed.

In Sec.~\ref{s6}, we recall a notion of a  $p$-adic {\it quasi-asymptotics\/}
from our papers~\cite{Kh-Sh1},~\cite{Kh-Sh2}.

In Sec.~\ref{s7}, a few multidimensional Tauberian type
theorems (Theorems~\ref{th5}--~\ref{th10}, Corollary~\ref{cor6})
for distributions are proved.
Theorem~\ref{th5} and Corollary~\ref{cor6} are related to the Fourier
transform and hold for distributions from ${\cD}'(\bQ_p^n)$.
Theorems~\ref{th7}--~\ref{th9} are related to the fractional operators
and hold for distributions from the Lizorkin spaces $\Phi_{\times}'(\bQ_p^n)$
and $\Phi'(\bQ_p^n)$. Theorem~\ref{th10} is  related to the
pseudo-differential operator (\ref{64.3}) in the Lizorkin space
$\Phi'(\bQ_p^n)$.

\section{$p$-Adic distributions.}\label{s2}

We shall use the notations and results from~\cite{Vl-V-Z}.
We denote by $\bZ$ the sets of integers numbers.
Recall that the field $\bQ_p$ of $p$-adic numbers is defined as the
completion of the field of rational numbers $\bQ$ with respect to the
non-Archimedean $p$-adic norm $|\cdot|_p$. This norm is defined as
follows: $|0|_p=0$; if an arbitrary rational number $x\ne 0$ is
represented as $x=p^{\gamma}\frac{m}{n}$, where $\gamma=\gamma(x)\in \bZ$,
and $m$ and $n$ are not divisible by $p$, then $|x|_p=p^{-\gamma}$.
This norm in $\bQ_p$ satisfies the strong triangle inequality
$|x+y|_p\le \max(|x|_p,|y|_p)$.

Denote by $\bQ_p^{*}=\bQ_p\setminus\{0\}$ the multiplicative group
of the field $\bQ_p$.
The space $\bQ_p^n=\bQ_p\times\cdots\times\bQ_p$ consists of points
$x=(x_1,\dots,x_n)$, where $x_j \in \bQ_p$, $j=1,2\dots,n$, \ $n\ge 2$.
The $p$-adic norm on $\bQ_p^n$ is
\begin{equation}
\label{8}
|x|_p=\max_{1 \le j \le n}|x_j|_p, \quad x\in \bQ_p^n.
\end{equation}

Denote by $B_{\gamma}^n(a)=\{x: |x-a|_p \le p^{\gamma}\},$ the ball
of radius $p^{\gamma}$ with the center at a point $a=(a_1,\dots,a_n)\in \bQ_p^n$
and $B_{\gamma}^n(0)=B_{\gamma}^n$, \ $\gamma \in \bZ$.
Here
\begin{equation}
\label{9}
B_{\gamma}^n(a)=B_{\gamma}(a_1)\times\cdots\times B_{\gamma}(a_n),
\end{equation}
where $B_{\gamma}(a_j)=\{x_j: |x_j-a_j|_p \le p^{\gamma}\}$ is a disc
of radius $p^{\gamma}$ with the center at a point $a_j\in \bQ_p$,
$j=1,2\dots,n$.

On $\bQ_p$ there exists the Haar measure, i.e., a positive measure $dx$
invariant under shifts, $d(x+a)=dx$, and normalized
by the equality $\int_{|\xi|_p\le 1}\,dx=1$.
The invariant measure $dx$ on the field $\bQ_p$ is extended to an
invariant measure $d^n x=dx_1\cdots dx_n$ on $\bQ_p^n$ in the standard way.

A complex-valued function $f$ defined on $\bQ_p^n$ is called
{\it locally-constant} if for any $x\in \bQ_p^n$ there exists
an integer $l(x)\in \bZ$ such that
$$
f(x+x')=f(x), \quad x'\in B_{l(x)}^n.
$$

Denote by ${\cE}(\bQ_p^n)$ and ${\cD}(\bQ_p^n)$ the
linear spaces of locally-constant $\bC$-valued functions on $\bQ_p^n$
and locally-constant $\bC$-valued functions with compact supports
(so-called test functions), respectively; ${\cD}={\cD}(\bQ_p)$,
${\cE}={\cE}(\bQ_p)$.
If $\varphi \in {\cD}(\bQ_p^n)$, according to Lemma~1 from~\cite[VI.1.]{Vl-V-Z},
there exists $l\in \bZ$, such that
$$
\varphi(x+x')=\varphi(x), \quad x'\in B_l^n, \quad x\in \bQ_p^n.
$$
The largest of such numbers $l=l(\varphi)$ is called the
{\it parameter of constancy} of the function $\varphi$.

Let us denote by ${\cD}^l_N(\bQ_p^n)$ the finite-dimensional space of
test functions from ${\cD}(\bQ_p^n)$ having supports in the ball $B_N^n$
and with parameters of constancy $\ge l$.
Any function $\varphi \in {\cD}^l_N(\bQ_p^n)$ is represented in the
following form
\begin{equation}
\label{9.4}
\varphi(x)=\sum_{\nu=1}^{p^{n(N-l)}}
\varphi(b^{\nu})\Delta_{l}(x_1-b_1^{\nu})\cdots\Delta_{l}(x_n-b_n^{\nu}),
\quad x\in \bQ_p^n,
\end{equation}
where $\Delta_{\gamma}(x_j-b_j^{\nu})$ is the characteristic function
of the ball $B_{l_j}(b_j^{\nu})$, and the points
$b^{\nu}=(b_1^{\nu},\dots b_n^{\nu})\in B_N^n$ do not depend on
$\varphi$~\cite[VI,(5.2')]{Vl-V-Z}

Denote by ${\cD}'(\bQ_p^n)$ the set of all linear functionals
(distributions) on ${\cD}(\bQ_p^n)$. It follows from~\cite[VI.3.]{Vl-V-Z}
that any linear functional $f$ is continuous on ${\cD}(\bQ_p^n)$.

Let us introduce in ${\cD}(\bQ_p^n)$ a {\it canonical
$\delta$-sequence} $\delta_k(x)\stackrel{def}{=}p^{nk}\Omega(p^k|x|_p)$,
and a {\it canonical $1$-sequence}
$\Delta_k(x)\stackrel{def}{=}\Omega(p^{-k}|x|_p)$, $k \in \bZ$, \
$x\in \bQ_p^n$, where
\begin{equation}
\label{10}
\Omega(t)=\left\{
\begin{array}{lcr}
1, &&\quad 0 \le t \le 1, \\
0, &&\quad t>1. \\
\end{array}
\right.
\end{equation}
Here $\Delta_k(x)$ is the characteristic function of the ball $B_{k}^n$.
It is clear~\cite[VI.3., VII.1.]{Vl-V-Z} that
$\delta_k \to \delta$, $k \to \infty$ in ${\cD}'(\bQ_p^n)$
and $\Delta_k \to 1$, $k \to \infty$ in ${\cE}(\bQ_p^n)$.

The convolution $f*g$ for distributions $f,g\in{\cD}'(\bQ_p^n)$ is
defined (see~\cite[VII.1.]{Vl-V-Z}) as
\begin{equation}
\label{11}
\langle f*g,\varphi\rangle
=\lim_{k\to \infty}\langle f(x)\times g(y),\Delta_k(x)\varphi(x+y)\rangle
\end{equation}
if the limit exists for all $\varphi\in {\cD}(\bQ_p^n)$,
where $f(x)\times g(y)$ is the direct product of distributions.

The Fourier transform of $\varphi\in {\cD}(\bQ_p^n)$ is defined by the
formula
$$
F[\varphi](\xi)=\int_{\bQ_p^n}\chi_p(\xi\cdot x)\varphi(x)\,d^nx,
\quad \xi \in \bQ_p^n,
$$
where $\chi_p(\xi\cdot x)=\chi_p(\xi_1 x_1)\cdots \chi_p(\xi_n x_n)
=e^{2\pi i\sum_{j=1}^{n}\{\xi_j x_j\}_p}$, \
$\xi\cdot x$ is the scalar product of vectors, and the function
$\chi_p(\xi_j x_j)=e^{2\pi i\{\xi_j x_j\}_p}$ for every fixed
$\xi_j \in \bQ_p$ is an additive character of the field $\bQ_p$, \
$\{\xi_j x_j\}_p$ is the fractional part of a number $\xi_j x_j$, \
$j=1,\dots,n$~\cite[VII.2.,3.]{Vl-V-Z}. It is known that the
Fourier transform is a linear isomorphism ${\cD}(\bQ_p^n)$ into
${\cD}(\bQ_p^n)$.
Moreover, according to~\cite[Lemma~A.]{Taib1},~\cite[III,(3.2)]{Taib3},
~\cite[VII.2.]{Vl-V-Z},
\begin{equation}
\label{12}
\varphi(x) \in {\cD}^l_N(\bQ_p^n) \quad \text{iff} \quad
F\big[\varphi(x)\big](\xi) \in {\cD}^{-N}_{-l}(\bQ_p^n).
\end{equation}
We define the Fourier transform $F[f]$ of a distribution
$f\in {\cD}'(\bQ_p^n)$ by the relation~\cite[VII.3.]{Vl-V-Z}
\begin{equation}
\label{13}
\langle F[f],\varphi\rangle=\langle f,F[\varphi]\rangle,
\quad \forall \, \varphi\in {\cD}(\bQ_p^n).
\end{equation}

Let $A$ be a matrix and $b\in \bQ_p^n$. Then for a distribution
$f\in{\cD}'(\bQ_p^n)$ the following relation holds~\cite[VII,(3.3)]{Vl-V-Z}:
\begin{equation}
\label{14}
F[f(Ax+b)](\xi)
=|\det{A}|_p^{-1}\chi_p\big(-A^{-1}b\cdot \xi\big)F[f(x)]\big(A^{-1}\xi\big),
\quad \det{A} \ne 0.
\end{equation}
In particular, if $f\in{\cD}'(\bQ_p)$, \ $a\in \bQ_p^{*}$, \
$b\in \bQ_p$ then
$$
F[f(ax+b)](\xi)
=|a|_p^{-1}\chi_p\Big(-\frac{b}{a}\xi\Big)F[f(x)]\Big(\frac{\xi}{a}\Big).
$$

According to~\cite[IV,(3.1)]{Vl-V-Z},
\begin{equation}
\label{14.1}
F[\Delta_{k}](x)=\delta_{k}(x), \quad k\in \bZ, \qquad x \in \bQ_p^n.
\end{equation}
In particular, $F[\Omega](x)=\Omega(x)$.

If for distributions $f,g\in {\cD}'(\bQ_p^n)$ a convolution $f*g$
exists then~\cite[VII,(5.4)]{Vl-V-Z}
\begin{equation}
\label{15}
F[f*g]=F[f]F[g].
\end{equation}

It is well known (see, e.g.,~\cite[III.2.]{Vl-V-Z}) that any
{\it multiplicative character\/} $\pi$ of the field $\bQ_p$
can be represented as
\begin{equation}
\label{16}
\pi(x)\stackrel{def}{=}\pi_{\alpha}(x)=|x|_p^{\alpha-1}\pi_{1}(x),
\quad x \in \bQ_p,
\end{equation}
where $\pi(p)=p^{1-\alpha}$ and $\pi_{1}(x)$ is a
{\it normed multiplicative character\/} such that
\begin{equation}
\label{16.1}
\pi_1(x)=\pi_{1}(|x|_px), \quad \pi_1(p)=\pi_1(1)=1, \quad
|\pi_1(x)|=1.
\end{equation}
We denote $\pi_{0}=|x|_p^{-1}$.

\begin{Definition}
\label{de1} \rm
Let $\pi_{\alpha}$ be a multiplicative character of the
field $\bQ_p$.

{(a)} (~\cite[Ch.II,\S 2.3.]{G-Gr-P},~\cite[VIII.1.]{Vl-V-Z})
A distribution $f \in {\cD}'(\bQ_p)$ is called
{\it homogeneous} of degree $\pi_{\alpha}$ if for all
$\varphi \in {\cD}(\bQ_p)$ and $t \in \bQ_p^*$ we have the relation
$$
\Bigl\langle f,\varphi\Big(\frac{x}{t}\Big) \Bigr\rangle
=\pi_{\alpha}(t)|t|_p \langle f,\varphi \rangle,
$$
i.e., $f(tx)=\pi_{\alpha}(t)f(x)$, $t \in \bQ_p^{*}$.

{(b)} We say that a distribution $f \in {\cD}'(\bQ_p^n)$ is
{\it homogeneous} of degree $\pi_{\alpha}$ if for all $t \in \bQ_p^*$
we have
\begin{equation}
\label{17}
f(tx)=f(tx_1,\dots,tx_n)=\pi_{\alpha}(t)f(x),
\quad x=(x_1,\dots,x_n)\in \bQ_p^{n}.
\end{equation}
A {\it homogeneous} distribution of degree $\pi_{\alpha}(x)=|x|_p^{\alpha-1}$
($\alpha \ne 0$) is called homogeneous of degree~$\alpha-1$.
\end{Definition}

For every multiplicative character $\pi_{\alpha}(x)\ne \pi_{0}=|x|_p^{-1}$,
$x\ne 0$ a {\it homogeneous\/} distribution $\pi_{\alpha}\in {\cD}'(\bQ_p)$
of degree $\pi_{\alpha}(x)$ is defined by~\cite[VIII,(1.6)]{Vl-V-Z}
$$
\langle \pi_{\alpha},\varphi \rangle
=\int_{B_0}|x|_p^{\alpha-1}\pi_1(x)\big(\varphi(x)-\varphi(0)\big)\,dx
\qquad\qquad\qquad\qquad
$$
\begin{equation}
\label{24}
\qquad\qquad
+\int_{\bQ_p\setminus B_0}|x|_p^{\alpha-1}\pi_1(x)\varphi(x)\,dx
+\varphi(0)I_0(\alpha),
\end{equation}
for all $\varphi\in {\cD}(\bQ_p)$, where
$$
I_0(\alpha)=\int_{B_0}|x|_p^{\alpha-1}\pi_1(x)\,dx
=\left\{
\begin{array}{rcl}
0, \quad \pi_1(x) &\not\equiv& 1, \\
\frac{1-p^{-1}}{1-p^{-\alpha}}, \quad \pi_1(x) &\equiv& 1. \\
\end{array}
\right.
$$
$\alpha \ne \mu_j=\frac{2\pi i}{\ln p}j$, \ $j\in \bZ$.

\begin{Definition}
\label{de1.1} \rm
{(a)} (~\cite{Al-Kh-Sh1}~\cite{Al-Kh-Sh2}) A distribution
$f_m\in {\cD}'(\bQ_p)$ is said to be {\it associated
homogeneous {\rm(}in the wide sense{\rm)}\/} of
degree~$\pi_{\alpha}$ and order~$m$, \ $m \in \bN_{0}$, if
$$
\Bigl\langle f_m,\varphi\Big(\frac{x}{t}\Big)\Bigr\rangle
=\pi_{\alpha}(t)|t|_p \langle f_m,\varphi \rangle
+\sum_{j=1}^{m}\pi_{\alpha}(t)|t|_p\log_p^j|t|_p
\langle f_{m-j},\varphi \rangle
$$
for all $\varphi \in {\cD(\bQ_p)}$ and $t \in \bQ_p^*$, where
$f_{m-j}\in {\cD}'(\bQ_p)$ is an associated homogeneous distribution
of degree~$\pi_{\alpha}$ and order $m-j$, \ $j=1,2,\dots,m$, i.e.,
$$
f_m(tx)=\pi_{\alpha}(t)f_m(x)
+\sum_{j=1}^{m}\pi_{\alpha}(t)\log_p^j|t|_pf_{m-j}(x), \quad t \in \bQ_p^*.
$$
If $m=0$ we set that the above sum is empty.

{(b)} We say that a distribution
$f \in {\cD}'(\bQ_p^n)$ is {\it associated homogeneous
{\rm(}in the wide sense{\rm)}\/} of degree $\pi_{\alpha}$ and order~$m$, \
$m \in \bN_{0}$, if for all $t \in \bQ_p^*$ we have
\begin{equation}
\label{18}
f_m(tx)=f_m(tx_1,\dots,tx_n)=\pi_{\alpha}(t)f_m(x)
+\sum_{j=1}^{m}\pi_{\alpha}(t)\log_p^j|t|_pf_{m-j}(x),
\end{equation}
where $f_{m-j}\in {\cD}'(\bQ_p^n)$ is an associated homogeneous
distribution of degree~$\pi_{\alpha}$ and order $m-j$, \ $j=1,2,\dots,m$.

An {\it associated homogeneous {\rm(}in the wide sense{\rm)}\/} distribution
of degree $\pi_{\alpha}(t)=|t|_p^{\alpha-1}$ and order~$m$ is called
{\it associated homogeneous} of degree~$\alpha-1$ and order~$m$.

{(c)} Associated homogeneous distribution (in the wide sense) of order
$m=1$ is called {\it associated homogeneous} distribution (see~\cite{Ge-Sh}
and~\cite{Al-Kh-Sh1},~\cite{Al-Kh-Sh2}).
\end{Definition}

The theorem describing all one-dimensional
{\it associated homogeneous {\rm(}in the wide sense{\rm)}\/} distributions
was proved in~\cite{Al-Kh-Sh1},~\cite{Al-Kh-Sh2}.

According to~\cite{Al-Kh-Sh1},~\cite{Al-Kh-Sh2},~\cite[\S 3]{Al-Kh-Sh3},
an associated homogeneous distribution of
degree~$\pi_{\alpha}(x)=|x|_p^{\alpha-1}\pi_1(x) \ne |x|_p^{-1}$
and order $m$, \ $m\in \bN$ is defined as
$$
\langle \pi_{\alpha}(x)\log_p^m|x|_p,\varphi(x) \rangle
=\int_{B_0}|x|_p^{\alpha-1}\pi_1(x)\log_p^m|x|_p
\big(\varphi(x)-\varphi(0)\big)\,dx
$$
$$
+\int_{\bQ_p\setminus B_0}|x|_p^{\alpha-1}\pi_1(x)\log_p^m|x|_p\varphi(x)\,dx
\qquad\qquad
$$
\begin{equation}
\label{19.3}
\quad
+\varphi(0)\int_{B_0}|x|_p^{\alpha-1}\pi_1(x)\log_p^m|x|_p\,dx,
\quad \forall \, \varphi\in {\cD}(\bQ_p),
\end{equation}
where
$I_{0}(\alpha;m)=\int_{B_0}|x|_p^{\alpha-1}\pi_1(x)\log_p^m|x|_p\,dx
=\frac{d^m I_{0}(\alpha)}{d\alpha^m} \log_p^m e$.
In~\cite{Al-Kh-Sh1},~\cite{Al-Kh-Sh2},~\cite[\S 3]{Al-Kh-Sh3} an
associated homogeneous distribution of
degree $\pi_{0}(x)=|x|_p^{-1}$ and order $m$, $m\in \bN$ is defined as
$$
\Bigl\langle P\Big(\frac{\log_p^{m-1}|x|_p}{|x|_p}\Big),\varphi \Bigr\rangle
\qquad\qquad\qquad\qquad\qquad\qquad\qquad\qquad\qquad
$$
\begin{equation}
\label{19.5}
=\int_{B_0}\frac{\log_p^{m-1}|x|_p}{|x|_p}\big(\varphi(x)-\varphi(0)\big)\,dx
+\int_{\bQ_p\setminus B_0}\frac{\log_p^{m-1}|x|_p}{|x|_p}\varphi(x)\,dx,
\end{equation}
for all $\varphi\in {\cD}(\bQ_p)$.

The integrals
\begin{equation}
\label{25}
\Gamma_p(\alpha)\stackrel{def}{=}\Gamma_p(|x|_p^{\alpha-1})
=\int_{\bQ_p} |x|_p^{\alpha-1}\chi_p(x)\,dx
=\frac{1-p^{\alpha-1}}{1-p^{-\alpha}},
\end{equation}
\begin{equation}
\label{25.1}
\Gamma_p(\pi_{\alpha})\stackrel{def}{=}F[\pi_{\alpha}](1)
=\int_{\bQ_p} |x|_p^{\alpha-1}\pi_{1}(x)\chi_p(x)\,dx
\qquad\qquad\qquad
\end{equation}
are called $p$-adic $\Gamma$-{\it functions\/}
~\cite[VIII,(2.2),(2.17)]{Vl-V-Z}.

If $\pi_{\alpha}^1(x)$, $\pi_{\beta}^2(x)$ are multiplicative characters
then the following relation holds~\cite[VIII,(3.6)]{Vl-V-Z}:
\begin{equation}
\label{25.4}
\big(\pi_{\alpha}^1*\pi_{\beta}^2\big)(x)
={\cB}_p(\pi_{\alpha}^1,\pi_{\beta}^2)
|x|_p^{\alpha+\beta-1}\pi_{1}^1(x)\pi_{1}^2(x),
\quad x \in \bQ_p,
\end{equation}
where
\begin{equation}
\label{25.5}
{\cB}_p(\pi_{\alpha}^1,\pi_{\beta}^2)
=\frac{\Gamma_p(\pi_{\alpha}^1)\Gamma_p(\pi_{\beta}^2)}
{\Gamma_p(\pi_{\alpha}^1\pi_{\beta}^2|x|_p)},
\end{equation}
is the ${\cB}$-function.

The multidimensional homogeneous distribution
$|x|_p^{\alpha-n}\in {\cD}'(\bQ_p^n)$
of degree $\alpha-n$ is constructed as follows.
If $Re\,\alpha>0$ then the function $|x|_p^{\alpha-n}$
generates a regular functional
\begin{equation}
\label{63.0}
\langle |x|_p^{\alpha-n},\varphi \rangle
=\int_{\bQ_p^n}|x|_p^{\alpha-n}\varphi(x)\,d^nx,
\quad \forall \, \varphi\in {\cD}(\bQ_p^n).
\end{equation}
If  $Re\,\alpha \le 0$ this distribution is defined by means
of analytic continuation~\cite[(*)]{Taib1},~\cite[III,(4.3)]{Taib3},
~\cite[VIII,(4.2)]{Vl-V-Z}:
$$
\langle |x|_p^{\alpha-n},\varphi \rangle
=\int_{B_0^n}|x|_p^{\alpha-n}\big(\varphi(x)-\varphi(0)\big)\,d^nx
\qquad\qquad\qquad\qquad
$$
\begin{equation}
\label{63.1}
\qquad\qquad
+\int_{\bQ_p^n\setminus B_0^n}|x|_p^{\alpha-n}\varphi(x)\,d^nx
+\varphi(0)\frac{1-p^{-n}}{1-p^{-\alpha}},
\end{equation}
for all $\varphi\in {\cD}(\bQ_p^n)$, \ $\alpha\ne \mu_j=\frac{2\pi i}{\ln p}j$,
$j\in \bZ$, where $|x|_p$, \ $x\in \bQ_p^n$ is given by (\ref{8}).
The distribution $|x|_p^{\alpha-n}$ is an entire function of the complex
variable $\alpha$ everywhere except the points $\mu_j$, $j\in \bZ$,
where it has simple poles with residues $\frac{1-p^{-n}}{\log p}\delta(x)$.

Similarly to the one-dimensional case (\ref{19.5}), one can construct the
distribution $P(\frac{1}{|x|_p^{n}})$ called the principal value of the
function~$\frac{1}{|x|_p^{n}}$:
\begin{equation}
\label{63.1*}
\Bigl\langle P\Big(\frac{1}{|x|_p^{n}}\Big),\varphi \Bigr\rangle
=\int_{B_0^n}\frac{\varphi(x)-\varphi(0)}{|x|_p^{n}}\,d^nx
+\int_{\bQ_p^n\setminus B_0^n}\frac{\varphi(x)}{|x|_p^{n}}\,d^nx,
\end{equation}
for all $\varphi\in {\cD}(\bQ_p^n)$.
It is easy to show that this distribution is
{\it associated homogeneous\/} of degree $-n$ and order $1$
(see~\cite{Al-Kh-Sh1}~\cite{Al-Kh-Sh2}).

The Fourier transform of $|x|_p^{\alpha-n}$ is given by the
formula~\cite{Sm},~\cite[Theorem~2.]{Taib1},~\cite[III,Theorem~(4.5)]{Taib3},
~\cite[VIII,(4.3)]{Vl-V-Z}
\begin{equation}
\label{63.2}
F[|x|_p^{\alpha-n}]=\Gamma^{(n)}_p(\alpha)|\xi|_p^{-\alpha},
\quad \alpha \ne 0,\,n
\end{equation}
where the n-dimensional $\Gamma$-{\it function\/} $\Gamma^{(n)}_p(\alpha)$
is given by the following formulas~\cite{Sm},~\cite[Theorem~1.]{Taib1},
~\cite[III,Theorem~(4.2)]{Taib3},~\cite[VIII,(4.4)]{Vl-V-Z}:
$$
\Gamma_p^{(n)}(\alpha)\stackrel{def}{=}
\lim_{k\to\infty}
\int_{p^{-k}\le |x|_p\le p^{k}} |x|_p^{\alpha-n}\chi_p(u\cdot x)\,d^nx
\qquad\qquad\qquad\qquad
$$
\begin{equation}
\label{63.3}
\qquad\qquad
=\int_{\bQ_p^n} |x|_p^{\alpha-n}\chi_p(x_1)\,d^nx
=\frac{1-p^{\alpha-n}}{1-p^{-\alpha}}
\end{equation}
where $|u|_p=1$. Here $\Gamma_p^{(1)}(\alpha)=\Gamma_p(\alpha)$.

\section{The $p$-adic Lizorkin spaces}
\label{s3}

\subsection{The Lizorkin space of the first kind.}\label{s3.1}
Consider the subspaces of the space of test functions $\cD(\bQ_p^n)$
$$
\Psi_{\times}=\Psi_{\times}(\bQ_p^n)
\qquad\qquad\qquad\qquad\qquad\qquad\qquad\qquad\qquad\qquad\qquad
\qquad
$$
$$
\qquad
=\{\psi(\xi)\in \cD(\bQ_p^n):
\psi(\xi_1,\dots,\xi_{j-1},0,\xi_{j+1},\dots,\xi_{n})=0, \, j=1,2,\dots,n\}
$$
and
$$
\Phi_{\times}=\Phi_{\times}(\bQ_p^n)
=\{\phi: \phi=F[\psi], \, \psi\in \Psi_{\times}(\bQ_p^n)\}.
$$
Obviously, $\Psi_{\times}, \Phi_{\times} \ne \emptyset$.
Since the Fourier transform is a linear isomorphism
${\cD}(\bQ_p^n)$ into ${\cD}(\bQ_p^n)$, we have
$\Psi_{\times}, \, \Phi_{\times}\subset \cD(\bQ_p^n)$.
The space $\Phi_{\times}$ admits the following characterization:
$\phi\in \Phi_{\times}$ if and only if $\phi\in \cD(\bQ_p^n)$ and
\begin{equation}
\label{50}
\int_{\bQ_p}\phi(x_1,\dots,x_{j-1},x_{j},x_{j+1},\dots,x_{n})\,dx_j=0,
\quad j=1,2,\dots,n.
\end{equation}

The space $\Phi_{\times}$ is called the $p$-adic {\it Lizorkin
space of test functions of the first kind\/}. By analogy with
the $\bC$-case~\cite[2.2.]{Sam3},~\cite[\S 25.1.]{Sam-Kil-Mar},
$\Phi_{\times}$ can be equipped with the topology of the space
$\cD(\bQ_p^n)$ which makes $\Phi_{\times}$ a complete space.
The space $\Phi_{\times}'=\Phi_{\times}'(\bQ_p^n)$ is called
the $p$-adic {\it Lizorkin space of distributions of the first kind\/}.

Let $\Psi^{\perp}_{\times}(\bQ_p^n)
=\{f\in \cD'(\bQ_p^n): \langle f,\psi\rangle=0,
\, \forall \, \psi\in \Psi_{\times}\}$, i.e., $\Psi^{\perp}_{\times}(\bQ_p^n)$
be the set of functionals from $\cD'(\bQ_p^n)$ concentrated on
the set $\cup_{j=1}^n\{x\in \bQ_p^n: x_j=0\}$.
Let $\Phi^{\perp}_{\times}(\bQ_p^n)=\{f\in \cD'(\bQ_p^n):
\langle f,\phi\rangle=0, \, \forall \, \phi\in \Phi_{\times}\}$.
Thus $\Phi^{\perp}_{\times}$ and $\Psi^{\perp}_{\times}$ are subspaces
of functionals in $\cD'$ orthogonal to $\Phi_{\times}$ and $\Psi_{\times}$,
respectively. It is clear that the set $\Psi^{\perp}_{\times}$ consists
of linear combinations of functionals of the form
$f(\xi_1\dots,\widehat{\xi_j},\dots,\xi_n)$, $j=1,2,\dots,n$,
where the hat \ $\widehat{\,}$ \ over $\xi_j$ denotes deletion
of the corresponding variable from the vector $\xi=(\xi_1\dots,\xi_n)$.
The set $\Phi^{\perp}_{\times}$ consists of linear combinations
of functionals of the form
$g(x_1\dots,\widehat{x_j},\dots,x_n)\times\delta(x_j)$, $j=1,2,\dots,n$.

\begin{Proposition}
\label{pr1}
The spaces of linear and continuous functionals $\Phi'_{\times}$
and $\Psi'_{\times}$ can be identified with the quotient spaces
$$
\Phi'_{\times}=\cD'/\Phi^{\perp}_{\times}, \qquad
\Psi'_{\times}=\cD'/\Psi^{\perp}_{\times}
$$
modulo the subspaces $\Phi^{\perp}_{\times}$ and $\Psi^{\perp}_{\times}$,
respectively.
\end{Proposition}

\begin{proof}
This proposition can be proved in the same way as~\cite[Proposition~2.5.]{Sam3}.
It follows from the well-known assertion:
if $E$ is a topological vector space with a closed subspace $M$ then $E'$
can be identified with the quotient space $M'=E'/M^{\perp}$, where
$M^{\perp}=\{f\in E': \langle f,\varphi\rangle=0,\, \forall \, \varphi\in M\}$.
\end{proof}

Analogously to (\ref{13}), we define the Fourier transform of
distributions $f\in \Phi_{\times}'(\bQ_p^n)$ and $g\in \Psi_{\times}'(\bQ_p^n)$
by the relations:
\begin{equation}
\label{51}
\begin{array}{rcl}
\displaystyle
\langle F[f],\psi\rangle=\langle f,F[\psi]\rangle,
&& \forall \, \psi\in \Psi_{\times}(\bQ_p^n), \\
\displaystyle
\langle F[g],\phi\rangle=\langle g,F[\phi]\rangle,
&& \forall \, \phi\in \Phi_{\times}(\bQ_p^n). \\
\end{array}
\end{equation}
By definition, $F[\Phi_{\times}(\bQ_p^n)]=\Psi_{\times}(\bQ_p^n)$ and
$F[\Psi_{\times}(\bQ_p^n)]=\Phi_{\times}(\bQ_p^n)$, i.e., (\ref{51})
give well defined objects. Moreover,
$F[\Phi_{\times}'(\bQ_p^n)]=\Psi_{\times}'(\bQ_p^n)$ and
$F[\Psi_{\times}'(\bQ_p^n)]=\Phi_{\times}'(\bQ_p^n)$,

\subsection{The Lizorkin space of the second kind.}\label{s3.2}
Now we consider the spaces
$$
\Psi=\Psi(\bQ_p^n)
=\{\psi(\xi)\in \cD(\bQ_p^n): \psi(0)=0\}
$$
and
$$
\Phi=\Phi(\bQ_p^n)=\{\phi: \phi=F[\psi], \, \psi\in \Psi(\bQ_p^n)\}.
$$
Here $\Psi, \Phi\subset \cD(\bQ_p^n)$.
The space $\Phi(\bQ_p^n)$ is called the $p$-adic {\it Lizorkin space of
test functions of the second kind\/}. Similarly to $\Phi_{\times}$,
the space $\Phi$ can be equipped with the topology of the space
$\cD(\bQ_p^n)$ which makes $\Phi$ a complete space.

Since the Fourier transform is a linear isomorphism ${\cD}(\bQ_p^n)$ into
${\cD}(\bQ_p^n)$, in view of (\ref{12}) the following lemma holds.
\begin{Lemma}
\label{lem1}
{\rm (a)} $\phi\in \Phi(\bQ_p^n)$ iff $\phi\in \cD(\bQ_p^n)$ and
\begin{equation}
\label{54}
\int_{\bQ_p^n}\phi(x)\,d^nx=0.
\end{equation}

{\rm (b)} $\phi \in {\cD}^l_N(\bQ_p^n)\cap\Phi(\bQ_p^n)$, i.e.,
$$
\int_{B^n_{N}}\phi(x)\,d^nx=0,
$$
iff $\psi=F^{-1}[\phi]\in {\cD}^{-N}_{-l}(\bQ_p^n)\cap\Psi(\bQ_p^n)$,
i.e.,
$$
\psi(\xi)=0, \qquad \xi \in B^n_{-N}.
$$
\end{Lemma}

In fact, for $n=1$, this lemma was proved in~\cite[IX.2.]{Vl-V-Z}.
Unlike the $\bC$-case situation (\ref{5}), (\ref{6}),
any function $\psi(\xi)\in \Phi$ is equal to zero not only at $\xi=0$
but in a ball $B^n \ni 0$, as well.

It follows from (\ref{54}) that the space $\Phi(\bQ_p^n)$ does not
contain real-valued functions everywhere different from zero.

Let $\Phi'=\Phi'(\bQ_p^n)$ denote the topological dual of the space
$\Phi(\bQ_p^n)$. We call it the $p$-adic {\it Lizorkin space of distributions
of the second kind\/}.

By $\Psi^{\perp}$ and $\Phi^{\perp}$ we denote the
subspaces of functionals in $\cD'$ orthogonal to $\Psi$ and
$\Phi$, respectively. Thus
$\Psi^{\perp}=\{f\in \cD'(\bQ_p^n): f=C\delta, \, C\in \bC\}$ and
$\Phi^{\perp}=\{f\in \cD'(\bQ_p^n): f=C, \, C\in \bC\}$.

\begin{Proposition}
\label{pr2}
$$
\Phi'=\cD'/\Phi^{\perp}, \qquad \Psi'=\cD'/\Psi^{\perp}.
$$
\end{Proposition}

This assertion is proved in the same way as Proposition~\ref{pr1}.

The space $\Phi'(\bQ_p^n)$ can be obtained from $\cD'(\bQ_p^n)$ by
``sifting out'' constants. Thus two distributions in $\cD'(\bQ_p^n)$
differing by a constant are indistinguishable as elements of $\Phi'(\bQ_p^n)$.

We define the Fourier transform of distributions $f\in \Phi'(\bQ_p^n)$
and $g\in \Psi'(\bQ_p^n)$ by an analog of formula (\ref{51}).
It is clear that $F[\Phi'(\bQ_p^n)]=\Psi'(\bQ_p^n)$
and $F[\Psi'(\bQ_p^n)]=\Phi'(\bQ_p^n)$,

Let $\Psi'_{M}(\bQ_p^n)$ be a class of multipliers in $\Psi(\bQ_p^n)$
and $\Phi'_{*}(\bQ_p^n)$ a class of convolutes in $\Phi(\bQ_p^n)$.
It is clear that a distribution $f\in \Psi'(\bQ_p^n)$ is a multiplier in
$\Psi(\bQ_p^n)$ if and only if $f\in \cE(\bQ_p^n\setminus\{0\})$.
Thus $\Phi'_{*}(\bQ_p^n)=F[\Psi'_{M}(\bQ_p^n)]$.
Since $\cE(\bQ_p^n)\subset \Psi'_M(\bQ_p^n)$, according to the
theorem from~\cite[VII.3.]{Vl-V-Z}, the class of all compactly supported
distributions from $f\in \cD'(\bQ_p^n)$ is a subset of $\Phi'_{*}(\bQ_p^n)$ .

\subsection{Density of the Lizorkin spaces in $\cL^{\rho}(\bQ_p^n)$.}\label{s3.3}
Repeating the proof of the assertions from~\cite{Sam2},~\cite[2.2.,2.4.]{Sam3}
practically word for word, we obtain the following $p$-adic analogs
of these assertions.

\begin{Lemma}
\label{lem2.1}
Let $g(\cdot)\in \cL^{1}(\bQ_p^n)$ and $f(\cdot)\in \cL^{\rho}(\bQ_p^n)$,
$1<\rho<\infty$. Then
$$
h_{t}(x)=\int_{\bQ_p^n}g(y)f(x-ty)\,d^ny
\qquad\qquad\qquad\qquad\qquad\qquad\qquad\qquad\qquad\quad
$$
\begin{equation}
\label{55}
=\frac{1}{|t|_p^n}
\int_{\bQ_p^n}g\Big(\frac{\xi}{t}\Big)f(x-\xi)\,d^n\xi
\stackrel{\cL^{\rho}}{\to}0,
\quad |t|_p \to \infty, \quad t\in \bQ_p^{*},
\quad x\in \bQ_p^{n}.
\end{equation}
\end{Lemma}

\begin{proof}
If $\rho=2$, taking into account the Parseval equality~\cite[VII,(4.4)]{Vl-V-Z},
formula (\ref{14}), and using the Riemann-Lebesgue lemma~\cite[VII.3.]{Vl-V-Z},
we have
\begin{equation}
\label{55.1}
||h_{t}||_{2}=||F[h_{t}]||_{2}=\bigg(\int_{\bQ_p^n}
\big|F[g](ty) \, F[f](y) \, \big|^2\,d^ny\bigg)^{\frac{1}{2}}\to 0,
\quad |t|_p \to \infty.
\end{equation}
Here the passage to the limit under the integral sign is justified
by the Lebesgue dominated theorem~\cite[IV.4]{Vl-V-Z}.

Let now $\rho \ne 2$. In view of the Young inequality~\cite[III,(1.7)]{Taib3},
$h_{t}(x)\in \cL^{\rho}(\bQ_p^n)$ and
\begin{equation}
\label{55.2}
||h_{t}||_{\rho} \le ||g||_{1} \, ||f||_{\rho},
\end{equation}
where the last estimate is uniform.

Clearly, it is sufficient to prove (\ref{55}) for $f\in \cD(\bQ_p^n)$.
Let $r>1$ be such that $\rho$ is located between $2$ and $r$. Using
the H\"older inequality and taking into account that $f\in \cD(\bQ_p^n)$,
we obtain
\begin{equation}
\label{55.3}
||h_{t}||_{\rho} \le ||h_{t}||_{r}^{1-\lambda} \, ||h_{t}||_{2}^{\lambda},
\end{equation}
where $\frac{1}{\rho}=\frac{1-\lambda}{r}+\frac{\lambda}{2}$
(i.e., $\lambda=\frac{2(\rho-r)}{\rho(2-r)}$). Since the lemma holds for
$\rho=2$, i.e., $||h_{t}||_{2}\to 0$, $|t|_p \to \infty$, by (\ref{55.1}),
(\ref{55.2}), we have
$$
||h_{t}||_{\rho} \le \big(||g||_{1} \, ||f||_{r}\big)^{1-\lambda} \,
||h_{t}||_{2}^{\lambda}\to 0, \quad |t|_p \to \infty,
\quad t\in \bQ_p^{*}.
$$
The lemma is thus proved.
\end{proof}

\begin{Lemma}
\label{lem2.1*}
Let $g(\cdot)\in \cL^{1}(\bQ_p^{n-m})$, $m\le n-1$ and
$f(\cdot)\in \cL^{\rho}(\bQ_p^n)$, \ $1<\rho<\infty$. Then
\begin{equation}
\label{55*}
h_{t}(x)=\int_{\bQ_p^{n-m}}g(y)f(x',x''-ty)\,d^{n-m}y
\stackrel{\cL^{\rho}}{\to}0,
\quad |t|_p \to \infty, \quad t\in \bQ_p^{*},
\end{equation}
where $x'=(x_1,\dots,x_m)\in \bQ_p^{m}$,
$x''=(x_{m+1},\dots,x_n)\in \bQ_p^{n-m}$, \ $1\le m \le n-1$.
\end{Lemma}

\begin{proof}
If $\rho=2$, just as above, using the Parseval
equality~\cite[VII,(4.4)]{Vl-V-Z}, and formula (\ref{14}), we have
\begin{equation}
\label{55.1*}
||h_{t}||_{2}=||F[h_{t}]||_{2}=\bigg(\int_{\bQ_p^n}
\big|F[g](ty'') \, F[f](y) \, \big|^2\,d^ny\bigg)^{\frac{1}{2}}\to 0,
\quad |t|_p \to \infty.
\end{equation}

Let now $\rho \ne 2$. In view of the Young inequality, we have the
uniform estimate
\begin{equation}
\label{55.2*}
||h_{t}||_{\cL^{\rho}(\bQ_p^n)} \le
||g||_{\cL^{1}(\bQ_p^{n-m})} \, ||f||_{\cL^{\rho}(\bQ_p^n)},
\end{equation}

Let $r>1$ be such that $\rho$ is located between $2$ and $r$.
Setting $f\in \cD(\bQ_p^n)$, using inequality (\ref{55.3}),
and taking into account that $||h_{t}||_{2}\to 0$, $|t|_p \to \infty$,
we obtain
$$
||h_{t}||_{\cL^{\rho}(\bQ_p^n)} \le
\big(||g||_{\cL^{1}(\bQ_p^{n-m})} \, ||f||_{\cL^{r}(\bQ_p^n)}\big)^{1-\lambda}
\, ||h_{t}||_{\cL^{2}(\bQ_p^n)}^{\lambda}\to 0, \quad |t|_p \to \infty.
$$
The lemma is thus proved.
\end{proof}

\begin{Lemma}
\label{lem2.2}
The space $\Phi(\bQ_p^n)$ is dense in $\cL^{\rho}(\bQ_p^n)$, $1<\rho<\infty$.
\end{Lemma}

\begin{proof}
Since $\cD(\bQ_p^n)$ is dense in $\cL^{\rho}(\bQ_p^n)$, $1<\rho<\infty$
(see~\cite[VI.2.]{Vl-V-Z}), it is sufficient to approximate the
function $\varphi\in \cD(\bQ_p^n)$ by functions
$\phi_{t}\in \Phi(\bQ_p^n)$ in the norm of $\cL^{\rho}(\bQ_p^n)$.

Consider a family of functions
$$
\psi_{t}(\xi)=(1-\Delta_{t}(\xi))F^{-1}[\varphi](\xi) \in \Psi(\bQ_p^n),
$$
where $\Delta_{t}(\xi)=\Omega(|t\xi|_p)$ is the characteristic function of
the ball $B_{\log_{p}|t|_p^{-1}}^n$,  $x\in \bQ_p^n$, \ $t\in \bQ_p^{*}$, \
the function $\Omega$ is defined by (\ref{10}).
In view of (\ref{15}), we have
$$
\phi_{t}(x)=F[\psi_{t}](x)=F[\big(1-\Delta_{t}(\xi)\big)](x)*\varphi(x)
\qquad\qquad\qquad\qquad
$$
$$
\qquad\qquad\qquad
=\delta(x)*\varphi(x)-F[\Delta_{t}(\xi)](x)*\varphi(x) \in \Phi(\bQ_p^n).
$$
According to (\ref{14.1}),
$F[\Delta_{t}(\xi)](x)=\frac{1}{|t|_p^{n}}\Omega\big(\frac{|x|_p}{|t|_p}\big)$,
i.e., the last relation can be rewritten as follows
$$
\phi_{t}(x)=\varphi(x)-\int_{\bQ_p^n}\Omega(|y|_p)\varphi(x-ty)d^ny.
$$

Applying Lemma~\ref{lem2.1} to the last relation, we see that
$||\phi_{t}-\varphi||_{\rho}\to 0$ as $|t|_p \to \infty$.
\end{proof}

\begin{Lemma}
\label{lem2.3}
The space $\Phi_{\times}(\bQ_p^n)$ is dense in $\cL^{\rho}(\bQ_p^n)$,
$1<\rho<\infty$.
\end{Lemma}

\begin{proof}
The proof of this lemma is based on the same calculations
as those carried out above.
In this case we set $\varphi\in \cD(\bQ_p^n)$ and
$$
\psi_{t}(\xi)=(1-\Delta_{t}(\xi_1)\cdots-\Delta_{t}(\xi_n))F^{-1}[\varphi](\xi)
\in \Psi_{\times}(\bQ_p^n),
$$
where $\Delta_{t}(\xi_j)=\Omega(|t\xi_j|_p)$ is the characteristic
function of the disc $B_{\log_{p}|t|_p^{-1}}$, $x_j\in \bQ_p$, \
$t\in \bQ_p^{*}$, \ $j=1,\dots,n$.
By (\ref{15}), we obtain
$$
\phi_{t}(x)
=\varphi(x)
-\big(\delta(x_2,\dots,x_n)\times F[\Delta_{t}(\xi_1)](x_1)\big)*\varphi(x)
\qquad\qquad\qquad\quad
$$
$$
\qquad\quad
\cdots
-\big(\delta(x_1,\dots,x_{n-1})\times F[\Delta_{t}(\xi_n)](x_n)\big)*\varphi(x)
\in \Phi_{\times}(\bQ_p^n).
$$
Since $F[\Delta_{t}(\xi_j)](x_j)
=\frac{1}{|t|_p}\Omega\big(\frac{|x_j|_p}{|t|_p}\big)$, $x_j\in \bQ_p$,
\ $j=1,\dots,n$, the last relation can be rewritten as
$$
\phi_{t}(x)=\varphi(x)
-\int_{\bQ_p}\Omega(|y_1|_p)\varphi(x_1-ty_1,x_2,\dots,x_n)dy_1
\qquad\qquad
$$
$$
\qquad\qquad
\cdots
-\int_{\bQ_p}\Omega(|y_n|_p)\varphi(x_1,\dots,x_{n-1},x_n-ty_n)dy_n.
$$

According to Lemma~\ref{lem2.1*},
$$
h_{j,t}(x)=\int_{\bQ_p}\Omega(|y_j|_p)
\varphi(x_1,\dots,x_{j-1},x_j-ty_j,x_{j+1},\dots,x_n)dy_j
\stackrel{\cL^{\rho}}{\to}0
$$
as $|t|_p \to \infty$, \ $j=1,\dots,n$. Thus
$||\phi_{t}-\varphi||_{\rho}
\le ||h_{1,t}||_{\rho}+\cdots+||h_{n,t}||_{\rho}\to 0$
as $|t|_p \to \infty$.
\end{proof}

For $n=1$ and $\rho=2$ the statements of Lemmas~\ref{lem2.2},~\ref{lem2.3}
coincide with the lemma from~\cite[IX.4.]{Vl-V-Z}

\section{Fractional operators}
\label{s4}

\subsection{The Vladimirov operator.}\label{s4.1}
Let us introduce a distribution from the space $\cD'(\bQ_p)$
\begin{equation}
\label{56}
f_{\alpha}(z)=\frac{|z|_p^{\alpha-1}}{\Gamma_p(\alpha)},
\quad \alpha \ne \mu_j, \quad \alpha \ne 1+\mu_j,
\quad z \in \bQ_p,
\end{equation}
called the {\it Riesz kernel\/}~\cite[VIII.2.]{Vl-V-Z}, where
$\mu_j=\frac{2\pi i}{\ln p}j$, $j\in \bZ$, \ $|z|_p^{\alpha-1}$
is a homogeneous distribution of degree~$\pi_{\alpha}(z)=|z|_p^{\alpha-1}$
defined by (\ref{24}), the $\Gamma$-function $\Gamma_p(\alpha)$ is given
by (\ref{25}).
The distribution $f_{\alpha}(z)$ is an entire function of the complex
variable $\alpha$ and has simple poles at the points $\alpha=\mu_j$,
$\alpha=1+\mu_j$, $j\in \bZ$.

According to~\cite[VIII,(2.20)]{Vl-V-Z}, we define $f_{0}(\cdot)$
as a distribution from $\cD'(\bQ_p)$:
\begin{equation}
\label{56.1}
f_{0}(z)\stackrel{def}{=}\lim_{\alpha \to 0}f_{\alpha}(z)=\delta(z),
\quad z \in \bQ_p,
\end{equation}
where the limit is understood in the weak sense.

Using~\cite[IX,(2.3)]{Vl-V-Z}, we define $f_{1}(\cdot)$
as a distribution from $\Phi'(\bQ_p)$:
\begin{equation}
\label{56.2}
f_{1}(z)\stackrel{def}{=}\lim_{\alpha \to 1}f_{\alpha}(z)
=-\frac{p-1}{\log p}\log|z|_p,
\quad z \in \bQ_p,
\end{equation}
where the limit is understood in the weak sense.

It is easy to see that if $\alpha \ne 1$ then the Riesz kernel
$f_{\alpha}(z)$ is a {\it homogeneous\/} distribution of
degree~$\alpha-1$, and if $\alpha=1$ then the Riesz kernel
is an {\it associated homogeneous\/} distribution of degree
$0$ and order $1$ (see Definitions~\ref{de1},~\ref{de1.1}).

It is well known that
\begin{equation}
\label{57.1}
f_{\alpha}(z)*f_{\beta}(z)=f_{\alpha+\beta}(z), \qquad
\alpha, \ \beta, \ \alpha+\beta \ne 1,
\end{equation}
in the sense of the space ${\cD}'(\bQ_p)$~\cite[VIII,(2.20),(3.8),(3.9)]{Vl-V-Z}.
Formulas (\ref{57.1}), (\ref{56.2}), i.e., in fact, results
of~\cite[IX.2]{Vl-V-Z}, imply that
\begin{equation}
\label{57.2}
f_{\alpha}(z)*f_{\beta}(z)=f_{\alpha+\beta}(z), \qquad
\alpha, \ \beta \in \bC,
\end{equation}
in the sense of distributions from ${\Phi}'(\bQ_p)$.

Let $\alpha=(\alpha_1,\dots,\alpha_n)$, $\alpha_j\in \bC$, $j=1,2,\dots$,
and $|\alpha|=\alpha_1+\cdots+\alpha_n$. We denote by
\begin{equation}
\label{58}
f_{\alpha}(x)=f_{\alpha_1}(x_1)\times\cdots\times f_{\alpha_n}(x_n),
\end{equation}
the {\it multi-Riesz kernel\/}, where the one-dimensional Riesz kernel
$f_{\alpha_j}(x_j)$, $j=1,\dots,n$ is defined by (\ref{56})--(\ref{56.2}).

If $\alpha_j \ne 1$, $j=1,2,\dots$ then the Riesz kernel
$$
f_{\alpha}(x)=\frac{|x_1|_p^{\alpha_1-1}}{\Gamma_p(\alpha_1)}
\times\cdots\times \frac{|x_n|_p^{\alpha_n-1}}{\Gamma_p(\alpha_n)}
$$
is a {\it homogeneous\/} distribution of degree~$|\alpha|-n$
(see Definition~\ref{de1}.(b)).

If $\alpha_1=\cdots=\alpha_k=1$, \  $\alpha_{k+1},\cdots,\alpha_{n}\ne 1$
then
$$
f_{\alpha}(x)
=(-1)^k\frac{(p-1)^k}{\log^k p}\log|x_1|_p\times\cdots\times\log|x_k|_p
\qquad\qquad\qquad\qquad
$$
\begin{equation}
\label{58.1}
\qquad\qquad\qquad\quad
\times\frac{|x_{k+1}|_p^{\alpha_{k+1}-1}}{\Gamma_p(\alpha_{k+1})}
\times\cdots\times \frac{|x_n|_p^{\alpha_n-1}}{\Gamma_p(\alpha_n)}.
\end{equation}
Thus, if among all $\alpha_1,\dots,\alpha_n$ there are $k$ pieces such
that $=1$ and $n-k$ pieces such that $\ne 1$ then the Riesz kernel
$f_{\alpha}(x)$ is an {\it associated homogeneous\/} distribution of
degree~$|\alpha|-n$ and order $k$, \ $k=1,\dots,n$
(see Definition~\ref{de1.1}.(b)).

For example, if $n=2$ and $\alpha_1=\alpha_2=1$ then we have
$f_{(1,1)}(x_1,x_2)=\frac{(p-1)^2}{\log^2 p}\log|x_1|_p\log|x_2|_p$, \
$x=(x_1,x_2)\in \bQ_p^2$ and
$$
f_{(1,1)}(tx_1,tx_2)=\frac{(p-1)^2}{\log^2 p}
\Big(\log|x_1|_p\log|x_2|_p
\qquad\qquad\qquad\qquad\qquad\qquad
$$
$$
\qquad\qquad
+(\log|x_1|_p+\log|x_2|_p)\log|t|_p+\log^2|t|_p\Big), \quad t\in \bQ^*_p.
$$

Define the multi-dimensional Vladimirov operator of the first kind
$D^{\alpha}_{\times}: \phi(x) \to D^{\alpha}_{\times}\phi(x)$
as the convolution
$$
\Big(D^{\alpha}_{\times}\phi\Big)(x)\stackrel{def}{=}f_{-\alpha}(x)*\phi(x)
\qquad\qquad\qquad\qquad\qquad\qquad\qquad\qquad\qquad\quad
$$
\begin{equation}
\label{59}
=\langle f_{-\alpha_1}(x_1)\times\cdots\times f_{-\alpha_n}(x_n),
\phi(x-\xi)\rangle,
\quad x\in \bQ_p^n,
\end{equation}
where $\phi\in \Phi_{\times}(\bQ_p^n)$. Here
$D^{\alpha}_{\times}=D^{\alpha_1}_{x_1}\times\cdots\times D^{\alpha_n}_{x_n}$,
where $D^{\alpha_j}_{x_j}=f_{-\alpha_j}(x_j)*$, $j=1,2,\dots,n$.

It is known that in the general case,
$(D^{\alpha}_{\times}\varphi)(x) \not\in {\cD}(\bQ_p^n)$
for $\varphi \in {\cD}(\bQ_p^n)$~\cite[IX]{Vl-V-Z}, i.e.,
the Bruhat--Schwartz space $\cD(\bQ_p^n)$ is not invariant
under the operator $D^{\alpha}_{\times}$.

\begin{Lemma}
\label{lem4}
The Lizorkin space of the first kind $\Phi_{\times}(\bQ_p^n)$ is
invariant under the Vladimirov fractional operator
$D^{\alpha}_{\times}$. Moreover,
$$
D^{\alpha}_{\times}(\Phi_{\times}(\bQ_p^n))=\Phi_{\times}(\bQ_p^n).
$$
\end{Lemma}

\begin{proof}
Taking into account formula~\cite[VIII,(2.1)]{Vl-V-Z}
\begin{equation}
\label{60}
F[f_{\alpha_j}(x_j)](\xi)=|\xi_j|_p^{-\alpha_j},
\quad j=1,\dots,n
\end{equation}
and (\ref{59}), (\ref{15}), we see that
$$
F[D^{\alpha}_{\times}\phi](\xi)
=|\xi_1|_p^{-\alpha_1}\times\cdots\times|\xi_n|_p^{-\alpha_n}
F[\phi](\xi), \quad \phi \in \Phi_{\times}(\bQ_p^n).
$$
Since $F[\phi](\xi)\in \Psi_{\times}(\bQ_p^n)$ and
$|\xi_1|_p^{-\alpha_1}\times\cdots\times|\xi_n|_p^{-\alpha_n}
F[\phi](\xi)\in \Psi_{\times}(\bQ_p^n)$ for any
$\alpha=(\alpha_1,\dots,\alpha_n)\in \bC^n$
then $D^{\alpha}_{\times}\phi \in \Phi_{\times}(\bQ_p^n)$, i.e.,
$D^{\alpha}_{\times}(\Phi_{\times}(\bQ_p^n))\subset \Phi_{\times}(\bQ_p^n)$.
Moreover, any function from $\Psi_{\times}(\bQ_p^n)$ can be represented
as $\psi(\xi)=|\xi_1|_p^{\alpha_1}\times\cdots\times|\xi_n|_p^{\alpha_n}
\psi_1(\xi)$, $\psi_1 \in \Psi_{\times}(\bQ_p^n)$.
This implies that
$D^{\alpha}_{\times}(\Phi_{\times}(\bQ_p^n))=\Phi_{\times}(\bQ_p^n)$.
\end{proof}

In view of (\ref{60}), (\ref{15}), formula (\ref{59}) can be rewritten as
\begin{equation}
\label{61}
\big(D^{\alpha}_{\times}\phi\big)(x)
=F^{-1}\big[|\xi_1|_p^{\alpha_1}\times\cdots\times|\xi_n|_p^{\alpha_n}
F[\phi](\xi)\big](x),
\quad \phi \in \Phi_{\times}(\bQ_p^n).
\end{equation}

The operator $D^{\alpha}_{\times}=f_{-\alpha}(x)*$ is called
the operator of fractional partial differentiation of order
$|\alpha|$, for $\alpha_j>0$, $j=1,\dots,n$; the operator of
fractional partial integration of order $|\alpha|$, for
$\alpha_j<0$, $j=1,\dots,n$; for $\alpha_1=\cdots=\alpha_n=0$, \
$D^{0}_{\times}=\delta(x)*$ is the identity operator.

According to formulas (\ref{59}), (\ref{11}), we define the
Vladimirov fractional operator $D^{\alpha}_{\times}f$, \ $\alpha\in \bC^n$
of a distribution $f\in \Phi_{\times}'(\bQ_p^n)$ by the relation
\begin{equation}
\label{62}
\langle D^{\alpha}_{\times}f,\phi\rangle\stackrel{def}{=}
\langle f, D^{\alpha}_{\times}\phi\rangle,
\qquad \forall \, \phi\in \Phi_{\times}(\bQ_p^n).
\end{equation}

In view of (\ref{62}) and Lemma~\ref{lem4},
$D^{\alpha}_{\times}(\Phi_{\times}'(\bQ_p^n))=\Phi_{\times}'(\bQ_p^n)$.
Moreover, in view of (\ref{57.2}), the family of operators $D^{\alpha}_{\times}$,
$\alpha \in \bC^n$ forms an Abelian group: if $f \in \Phi'(\bQ_p^n)$ then
\begin{equation}
\label{63}
\begin{array}{rcl}
\displaystyle
D^{\alpha}_{\times}D^{\beta}_{\times}f&=&
D^{\beta}_{\times}D^{\alpha}_{\times}f=D^{\alpha+\beta}_{\times}f, \medskip \\
\displaystyle
D^{\alpha}_{\times}D^{-\alpha}_{\times}f
&=&f,
\qquad \alpha,\beta \in \bC^n, \\
\end{array}
\end{equation}
where $\alpha+\beta=(\alpha_1+\beta_1,\dots,\alpha_n+\beta_n)\in \bC^n$.

\begin{Example}
\label{ex1} \rm
If $\alpha_j>0$, $j=1,2,\dots$ then the fractional integration formula
for the delta function holds
$$
D^{-\alpha}_{\times}\delta(x)=\frac{|x_1|_p^{\alpha_1-1}}{\Gamma_p(\alpha_1)}
\times\cdots\times \frac{|x_n|_p^{\alpha_n-1}}{\Gamma_p(\alpha_n)}.
$$
\end{Example}

\subsection{The Taibleson operator.}\label{s4.2}
Let us introduce the distribution from ${\cD}'(\bQ_p^n)$
\begin{equation}
\label{63.4}
\kappa_{\alpha}(x)=\frac{|x|_p^{\alpha-n}}{\Gamma_p^{(n)}(\alpha)},
\quad \alpha \ne 0, \, n, \qquad x\in \bQ_p^n,
\end{equation}
called the  multidimensional {\it Riesz kernel\/}~\cite[\S2]{Taib1},
~\cite[III.4.]{Taib3}, where the function $|x|_p$, \ $x\in \bQ_p^n$
is given by (\ref{8}).
The Riesz kernel has a removable singularity at $\alpha=0$ and according
to~\cite[\S2]{Taib1},~\cite[III.4.]{Taib3},~\cite[VIII.2]{Vl-V-Z}, we have
$$
\langle \kappa_{\alpha}(x),\varphi(x)\rangle
=\frac{g_{\alpha}}{\Gamma_p^{(n)}(\alpha)}
+\frac{1-p^{-n}}{(1-p^{-\alpha})\Gamma_p^{(n)}(\alpha)}\varphi(0)
\qquad\qquad
$$
$$
\qquad\qquad
=g_{\alpha}\frac{1-p^{-\alpha}}{1-p^{\alpha-n}}
+\frac{1-p^{-n}}{1-p^{\alpha-n}}\varphi(0),
\quad \varphi\in {\cD}(\bQ_p^n),
$$
where $g_{\alpha}$ is an entire function in $\alpha$.
Passing to the limit in the above relation, we obtain
$$
\langle \kappa_{0}(x),\varphi(x)\rangle\stackrel{def}{=}
\lim_{\alpha\to 0}\langle \kappa_{\alpha}(x),\varphi(x)\rangle=\varphi(0),
\quad \forall \, \varphi\in {\cD}(\bQ_p^n).
$$
Thus we define $\kappa_{0}(\cdot)$ as a distribution from ${\cD}'(\bQ_p^n)$:
\begin{equation}
\label{63.5}
\kappa_{0}(x)\stackrel{def}{=}\lim_{\alpha\to 0}\kappa_{\alpha}(x)=\delta(x).
\end{equation}

Next, using (\ref{63.0}), (\ref{63.4}), and taking into account
(\ref{54}), we define $\kappa_{n}(\cdot)$ as a distribution from the
{\it Lizorkin space of distributions\/} $\Phi'(\bQ_p^n)$:
$$
\langle \kappa_{n}(x),\phi \rangle\stackrel{def}{=}
\lim_{\alpha \to n}\langle \kappa_{\alpha}(x),\phi \rangle
=\lim_{\alpha \to n}
\int_{\bQ_p^n}\frac{|x|_p^{\alpha-n}}{\Gamma^{(n)}_p(\alpha)}\phi(x)\,d^nx
\qquad\qquad
$$
$$
=-\lim_{\beta \to 0}\big(1-p^{-n-\beta}\big)
\int_{\bQ_p^n}\frac{|x|_p^{\beta}-1}{p^{\,\beta}-1}\phi(x)\,d^nx
\qquad
$$
$$
=-\frac{1-p^{-n}}{\log p}\int_{\bQ_p^n}\log|x|_p\phi(x)\,d^nx,
\quad \forall \, \phi\in \Phi(\bQ_p^n),
$$
where $|\alpha-n|\le 1$. Similarly to the one-dimensional
case~\cite[IX.2]{Vl-V-Z}, the passage to the limit under the integral
sign is justified by the Lebesgue dominated theorem~\cite[IV.4]{Vl-V-Z}.
Thus,
\begin{equation}
\label{63.7}
\kappa_{n}(x)\stackrel{def}{=}\lim_{\alpha \to n}\kappa_{\alpha}(x)
=-\frac{1-p^{-n}}{\log p}\log|x|_p.
\end{equation}

Thus the Riesz kernel $\kappa_{\alpha}(x)$ is well defined distribution
from the Lizorkin space $\Phi'(\bQ_p^n)$ for all $\alpha \in \bC$.

According to Definitions~\ref{de1}.(b) and~\ref{de1.1}.(b),
if $\alpha \ne n$ then $\kappa_{\alpha}(x)$ is a {\it homogeneous\/}
distribution of degree~$\alpha-n$, and if $\alpha=n$ then
$\kappa_{\alpha}(x)$ is an {\it associated homogeneous\/}
distribution of degree $0$ and order $1$.

With the help of (\ref{63.2}), (\ref{63.5}), we obtain the
formulas~\cite[(**)]{Taib1},~\cite[III,(4.6)]{Taib3},
~\cite[VIII,(4.9),(4.10)]{Vl-V-Z}:
\begin{equation}
\label{63.8}
\kappa_{\alpha}(x)*\kappa_{\beta}(x)=\kappa_{\alpha+\beta}(x),
\quad \alpha, \, \beta, \, \alpha+\beta \ne n,
\end{equation}
which holds in the sense of the space ${\cD}'(\bQ_p^n)$.
Taking into account formula (\ref{63.7}), it is easy to see that
\begin{equation}
\label{63.9}
\kappa_{\alpha}(x)*\kappa_{\beta}(x)=\kappa_{\alpha+\beta}(x),
\quad \alpha, \beta \in \bC,
\end{equation}
in the sense of the Lizorkin space $\Phi'(\bQ_p^n)$.

Define the multi-dimensional Taibleson operator in the Lizorkin space
$\phi\in \Phi(\bQ_p^n)$ as the convolution:
\begin{equation}
\label{59**}
\big(D^{\alpha}_{x}\phi\big)(x)\stackrel{def}{=}\kappa_{-\alpha}(x)*\phi(x)
=\langle \kappa_{-\alpha}(x),\phi(x-\xi)\rangle,
\quad x\in \bQ_p^n,
\end{equation}
$\phi\in \Phi(\bQ_p^n)$, \ $\alpha \in \bC$.

\begin{Lemma}
\label{lem4.1}
The Lizorkin space of the second kind $\Phi(\bQ_p^n)$ is
invariant under the Taibleson fractional operator
$D^{\alpha}_{x}$ and $D^{\alpha}_{x}(\Phi(\bQ_p^n))=\Phi(\bQ_p^n)$.
\end{Lemma}

\begin{proof}
The proof of Lemma~\ref{lem4.1} is carried out in the
same way as the proof of Lemma~\ref{lem4}.

In view of formula (\ref{63.2}),
$F[\kappa_{\alpha}(x)](\xi)=|\xi|_p^{-\alpha}$.
Consequently, using (\ref{15}), we have
$$
F[D^{\alpha}_{x}\phi](\xi)=|\xi|_p^{-\alpha}F[\phi](\xi),
\quad \phi \in \Phi(\bQ_p^n).
$$
Thus $F[\phi](\xi), |\xi_1|_p^{-\alpha}F[\phi](\xi)\in \Psi(\bQ_p^n)$, \
$\alpha\in \bC$ and $D^{\alpha}_{x}\phi \in \Phi(\bQ_p^n)$. That is
$D^{\alpha}_{x}(\Phi(\bQ_p^n))\subset \Phi(\bQ_p^n)$.
Since any function from $\Psi(\bQ_p^n)$ can be represented
as $\psi(\xi)=|\xi|_p^{\alpha}\psi_1(\xi)$, $\psi_1 \in \Psi(\bQ_p^n)$,
we have $D^{\alpha}_{x}(\Phi(\bQ_p^n))=\Phi(\bQ_p^n)$.
\end{proof}

In view of (\ref{63.2}), (\ref{15}), formula (\ref{59**}) can be
represented in the form
\begin{equation}
\label{61**}
\big(D^{\alpha}_{x}\phi\big)(x)
=F^{-1}\big[|\xi|^{\alpha}_pF[\phi](\xi)\big](x),
\quad \phi \in \Phi(\bQ_p^n).
\end{equation}

According to (\ref{59**}), (\ref{11}), we define
$D^{\alpha}f$ of a distribution $f\in \Phi'(\bQ_p^n)$ by the relation
\begin{equation}
\label{62**}
\langle D^{\alpha}_{x}f,\phi\rangle\stackrel{def}{=}
\langle f, D^{\alpha}_{x}\phi\rangle,
\qquad \forall \, \phi\in \Phi(\bQ_p^n).
\end{equation}

It is clear that $D^{\alpha}_{x}(\Phi'(\bQ_p^n))=\Phi'(\bQ_p^n)$
and the family of operators $D^{\alpha}_{x}$, $\alpha \in \bC$ have
group properties of the form (\ref{63}) on the space of distributions
$\Phi'(\bQ_p^n)$.

\begin{Example}
\label{ex2} \rm
If $\alpha>0$ then the fractional integration formula for the
delta function holds
$$
D^{-\alpha}\delta(x)=\frac{|x|_p^{\alpha-n}}{\Gamma_p^{(n)}(\alpha)}.
$$
\end{Example}

\begin{Remark}
\label{rem1} \rm
In~\cite[IX.5.]{Vl-V-Z}, the orthonormal complete basis in
${\cL}^2(\bQ_p)$ of eigenfunctions of Vladimirov's operator
$D^{\alpha}=f_{-\alpha}*$, $\alpha>0$ was constructed. Another
orthonormal complete basis in ${\cL}^2(\bQ_p)$ of eigenfunctions
of the operator $D^{\alpha}$, $\alpha>0$
\begin{equation}
\label{62.1}
\Theta_{\gamma j a}(x)=p^{-\gamma/2}\chi_p\big(p^{-1}j(p^{\gamma}x-a)x\big)
\Omega\big(|p^{\gamma}x-a|_p\big), \quad x\in \bQ_p,
\end{equation}
$\gamma\in \bZ$, $a\in I_p=\bQ_p/\bZ_p$, $j=1,2,\dots,p-1$, was
later constructed by S.~V.~Kozyrev in~\cite{Koz0}.
Here elements of the group $I_p=\bQ_p/\bZ_p$ can be
represented in the form
$$
a=p^{-\gamma}\big(a_{0}+a_{1}p^{1}+\cdots+a_{\gamma-1}p^{\gamma-1}\big),
\quad \gamma\in \bN,
$$
where $a_j=0,1,\dots,p-1$, \ $j=0,1,\dots,\gamma-1$.
Thus
\begin{equation}
\label{62.2}
D^{\alpha}\Theta_{\gamma j a}(x)=p^{\alpha(1-\gamma)}\Theta_{\gamma j a}(x),
\quad \alpha>0.
\end{equation}

Since, according to~\cite[IX,(5.7),(5.8)]{Vl-V-Z},~\cite{Koz0},
$\int_{\bQ_p}\Theta_{\gamma j a}(x)\,dx=0$, the eigenfunctions
$\Theta_{\gamma j a}(x)$ of Vladimirov's operator $D^{\alpha}$,
$\alpha>0$ belong to the Lizorkin space $\Phi(\bQ_p)$ (see Lemma~\ref{lem1}).
Since the Lizorkin space is invariant under the Vladimirov operator,
$\Theta_{\gamma j a}(x)$ {\it are also eigenfunctions\/} of Vladimirov's
operator $D^{\alpha}$ for $\alpha<0$, i.e., relation (\ref{62.2})
holds for any $\alpha\in \bC$.
\end{Remark}

\subsection{$p$-Adic Laplacians.}\label{s4.3}
By analogy with the ``$\bC$-case'' ~\cite{Sam3},~\cite{Sam-Kil-Mar},
and the $p$-adic case~\cite{Kh0},~\cite[X.1,Example~2]{Vl-V-Z},
using the fractional operators one can introduce the $p$-adic Laplacians.

The Laplacian of the first kind is an operator
$$
-\widehat\Delta f(x)\stackrel{def}{=}\sum_{k=1}^n\big(D^{2}_{x_k}f\big)(x),
\quad f\in \Phi'(\bQ_p^n)
$$
with the symbol $-\sum_{k=1}^n|\xi_j|_p^{2}$, $\xi_k\in \bQ_p$, $k=1,2,\dots,n$;
the Laplacian of the second kind is an operator
$$
-\Delta f(x)\stackrel{def}{=}\big(D^{2}_{x}f\big)(x),
\quad f\in \Phi'(\bQ_p^n).
$$
with the symbol $-|\xi|_p^{2}$, $\xi\in \bQ_p^n$.
Moreover, one can define powers of the Laplacian by the formula
$$
(-\Delta)^{\alpha/2}f(x)\stackrel{def}{=}\big(D^{\alpha}_{x}f\big)(x),
\quad f\in \Phi'(\bQ_p^n), \quad \alpha \in \bC.
$$

\section{Pseudo-differential operators and equations.}\label{s5}

Similarly to the representation (\ref{61**}), one can consider
a class of pseudo-differential operators in the Lizorkin space
of the test functions $\Phi(\bQ_p^n)$
$$
(A\phi)(x)=F^{-1}\big[\cA(\xi)\,F[\phi](\xi)\big](x)
\qquad\qquad\qquad\qquad\qquad\qquad\qquad
$$
\begin{equation}
\label{64.3}
=\int_{\bQ_p^n}\int_{\bQ_p^n}\chi_p\big((y-x)\cdot \xi\big)
\cA(\xi)\phi(y)\,d^n\xi\,d^ny,
\quad \phi \in \Phi(\bQ_p^n)
\end{equation}
with symbols $\cA(\xi)\in \cE(\bQ_p^n\setminus \{0\})$.

In view of Subsec.~\ref{s3.2}, functions
$F[\phi](\xi)$ and $\cA(\xi)F[\phi](\xi)$ belong to $\Psi(\bQ_p^n)$,
and, consequently, $(A\phi)(x)\in \Phi(\bQ_p^n)$. Thus the pseudo-differential
operators (\ref{64.3}) are well defined and the Lizorkin space
$\Phi(\bQ_p^n)$ is invariant under them.

If we define a conjugate pseudo-differential operator $A^{T}$ as
\begin{equation}
\label{64.5}
(A^{T}\phi)(x)=F^{-1}[\cA(-\xi)F[\phi](\xi)](x)
=\int_{\bQ_p^n}\chi_p(-x\cdot \xi)\cA(-\xi)F[\phi](\xi)\,d^n\xi
\end{equation}
then one can define operator $A$ in the Lizorkin space of distributions:
for $f \in \Phi'(\bQ_p^n)$ we have
\begin{equation}
\label{64.4}
\langle Af,\phi\rangle=\langle f,A^{T}\phi\rangle,
\qquad \forall \, \phi \in \Phi(\bQ_p^n).
\end{equation}
It is clear that
\begin{equation}
\label{64.3*}
Af=F^{-1}[\cA\,F[f]]\in \Phi'(\bQ_p^n),
\end{equation}
i.e., the Lizorkin space of distributions $\Phi'(\bQ_p^n)$
is invariant under pseudo-differential operators $A$.

If $A, B$ are pseudo-differential operators with symbols
$\cA(\xi), \cB(\xi)\in \cE(\bQ_p^n\setminus \{0\})$, respectively,
then the operator $AB$ is well defined and represented by the formula
$$
(AB)f=F^{-1}[\cA\cB\,F[f]]\in \Phi'(\bQ_p^n).
$$
If $\cA(\xi)\ne 0$, $\xi\in \bQ_p^n\setminus \{0\}$ then we define
the inverse pseudo-differential by the formula
$$
A^{-1}f=F^{-1}[\cA^{-1}\,F[f]], \quad f\in \Phi'(\bQ_p^n).
$$

Thus the family of pseudo-differential operators $A$ with symbols
$\cA(\xi)\ne 0$, $\xi\in \bQ_p^n\setminus \{0\}$ forms an Abelian group.

If the symbol $\cA(\xi)$ of the operator $A$ is an {\it associated homogeneous\/}
function then the operator $A$ is called an {\it associated homogeneous
pseudo-differential operator\/}.

According to formulas (\ref{61**}), (\ref{63.4})--(\ref{63.7}),
and Definitions~\ref{de1},~\ref{de1.1}
the operator $D^{\alpha}_{x}$, $\alpha\ne -n$ is
a {\it homogeneous\/} pseudo-differential operator of
degree~$\alpha$ with the symbol $\cA(\xi)=|\xi|_p^{\alpha}$
and $D^{-n}_{x}$ is a {\it homogeneous\/} pseudo-differential
operator of degree $-n$ and order $1$ with the symbol
$\cA(\xi)=P(|\xi|_p^{-n})$ (see (\ref{63.1*})).

Let us consider a pseudo-differential equation
\begin{equation}
\label{64.3**}
Af=g, \qquad g\in \Phi'(\bQ_p^n),
\end{equation}
where $A$ is a pseudo-differential operator (\ref{64.3}), $f$
is the desired distribution.

\begin{Theorem}
\label{th4.2}
If the symbol of a pseudo-differential operator $A$ is such that
$\cA(\xi)\ne 0$, $\xi\in \bQ_p^n\setminus \{0\}$ then the equation
{\rm (\ref{64.3**})} has the unique solution
$$
f(x)=F^{-1}\Big[\frac{F[g](\xi)}{\cA(\xi)}\Big](x)=(A^{-1}g)(x)\in \Phi'(\bQ_p^n).
$$
\end{Theorem}

\begin{proof}
Applying the Fourier transform to the left-hand and right-hand sides of
equation $Af=g$, in view of representation (\ref{64.3*}), we obtain that
$\cA(\xi)F[f](\xi)=F[g](\xi)$. Since according to Subsec.~\ref{s3.2},
$F[\Phi'(\bQ_p^n)]=\Psi'(\bQ_p^n)$, $F[\Psi'(\bQ_p^n)]=\Phi'(\bQ_p^n)$,
and $\cA(\xi)$ is a multiplier in $\Psi(\bQ_p^n)$, we have
$F[f](\xi)=\cA^{-1}(\xi)F[g](\xi)\in \Psi'(\bQ_p^n)$. Thus
$f(x)=F^{-1}[\cA^{-1}(\xi)F[g](\xi)](x)=(A^{-1}g)(x)\in \Phi'(\bQ_p^n)$
is a solution of the problem (\ref{64.3**}).

Now we study solutions of the homogeneous problem (\ref{64.1}).
Let $f\in {\cD}'(\bQ_p^n)$ and $Af=0$, i.e., according to (\ref{64.4}),
$\langle Af,\phi\rangle=\langle f,A^{T}\phi\rangle=0$, for all
$\phi\in \Phi(\bQ_p^n)$. Since $A^{T}(\Phi(\bQ_p^n))=\Phi(\bQ_p^n)$,
we have $\langle f,\phi\rangle=0$, for all $\phi\in \Phi(\bQ_p^n)$, and
consequently, $f\in \Phi^{\perp}$ (see Proposition~\ref{pr2}). Thus the
solutions of the homogeneous problem (\ref{64.3**}) are indistinguishable
as elements of the space $\Phi'(\bQ_p^n)$.
\end{proof}

Let $P_N(z)=\sum_{k=0}^Na_kz^{k}$ be a polynomial, where
$a_k\in \bC$ are constants.
Let us consider the equation
\begin{equation}
\label{64.1}
P_N\big(D^{\alpha}_{x}\big)f=g, \qquad g\in \Phi'(\bQ_p^n),
\end{equation}
where $\big(D^{\alpha}_{x}\big)^k\stackrel{def}{=}D^{\alpha k}_{x}$,
$\alpha \in \bC$ and $f$ is the desired distribution.

\begin{Theorem}
\label{th4}
If $P_N(z)\ne 0$ for all $z>0$ then equation {\rm (\ref{64.1})} has
the unique solution
\begin{equation}
\label{64.2}
f(x)=F^{-1}\Big[\frac{F[g](\xi)}{P_N\big(|\xi|_p^{\alpha}\big)}\Big](x)
\in \Phi'(\bQ_p^n).
\end{equation}
In particular, the unique solution of the equation
$$
D^{\alpha}_{x}f=g, \qquad g\in \Phi'(\bQ_p^n),
$$
is given by the formula
$f=D^{-\alpha}_{x}g\in \Phi'(\bQ_p^n)$.
\end{Theorem}

\begin{proof}
According to formulas (\ref{63.1})--(\ref{63.2}), (\ref{63.4})--(\ref{63.7}),
$$
F[\kappa_{\alpha}(x)]=|\xi|_p^{-\alpha}, \quad \alpha \in \bC
$$
in $\Phi'(\bQ_p^n)$. Consequently, applying the Fourier transform to
the left-hand and right-hand sides of relation (\ref{64.1}), we obtain
(\ref{64.2}). Here we must take into account the fact that
$\frac{1}{P_N(|\xi|_p^{\alpha})}$ is a multiplier in $\Psi(\bQ_p^n)$.
Thus (\ref{64.2}) is the solution of the problem (\ref{64.1}).

In view of the proof of Theorem~\ref{th4.2}, the homogeneous problem
(\ref{64.1}) has only a trivial solution.
\end{proof}

In a similar way we can prove the following theorem.
\begin{Theorem}
\label{th4.1}
If $P_N(z)\ne 0$ for all $z>0$ then
the equation
$$
P_N\big(D^{\alpha}_{\times}\big)f=g, \qquad g\in \Phi'_{\times}(\bQ_p^n),
$$
$\alpha \in \bC^n$ has the unique solution
$$
f(x)=F^{-1}\Big[\frac{F[g](\xi)}
{P_N\big(|\xi_1|_p^{\alpha_1}\cdots |\xi_n|_p^{\alpha_n}\big)}\Big](x)
\in \Phi'_{\times}(\bQ_p^n).
$$
In particular, the unique solution of the equation
$$
D^{\alpha}_{\times}f=g, \qquad g\in \Phi'_{\times}(\bQ_p^n),
$$
is given by formula $f=D^{-\alpha}_{\times}g\in \Phi'_{\times}(\bQ_p^n)$.
\end{Theorem}

Now we prove an analog of the statement for the Vladimirov
fractional operator~\cite[IX.1,Example~4]{Vl-V-Z}.

\begin{Proposition}
\label{pr4}
Let $A$ be a pseudo-differential operator with a symbol $\cA(\xi)$
and $0\ne z\in \bQ_p^n$. Then the additive character $\chi_p(z\cdot x)$
is an eigenfunction of the operator $A$ with the eigenvalue $\cA(-z)$,
i.e.,
$$
A\chi_p(z\cdot x)=\cA(-z)\chi_p(z\cdot x).
$$
\end{Proposition}

\begin{proof}
Since $F[\chi_p(z\cdot x)]=\delta(\xi+z)$, $z\ne 0$, we have
$\cA(\xi)\delta(\xi+z)=\cA(-z)\delta(\xi+z)$. Thus
$$
A\chi_p(z\cdot x)=F^{-1}[\cA(\xi)F[\chi_p(z\cdot x)](\xi)](x)
\qquad\qquad\qquad\qquad
$$
$$
\qquad\quad
=\cA(-z)F^{-1}[\delta(\xi+z)](x)=\cA(-z)\chi_p(z\cdot x).
$$
\end{proof}

\section{Distributional quasi-asymptotics.}
\label{s6}

We recall some facts from our papers~\cite{Kh-Sh1},~\cite{Kh-Sh2},
where we introduced the notion of the {\it quasi-asymptotics\/}~\cite{D-Zav1},
~\cite{Vl-D-Zav} adapted to the $p$-adic case.

\begin{Definition}
\label{de4.1} \rm
(~\cite{Kh-Sh1},~\cite{Kh-Sh2}) A continuous complex valued function
$\rho(z)$ on the multiplicative group $\bQ_p^*$ such that for any
$z\in \bQ_p^*$ the limit
$$
\lim_{|t|_p \to \infty}\frac{\rho(tz)}{\rho(t)}=C(z)
$$
exists is called an {\it automodel {\rm(}or regular varying{\rm)}\/}
function.
\end{Definition}

It is easy to see that the function $C(z)$ satisfies the functional
equation $C(ab)=C(a)C(b)$, \ $a,b\in \bQ_p^*$. According
to~\cite[Ch.II,\S 1.4.]{G-Gr-P},~\cite[III.2.]{Vl-V-Z}, the solution
of this equation is a multiplicative character $\pi_{\alpha}$ of the
field $\bQ_p$ defined by (\ref{16}), (\ref{16.1}), i.e.,
\begin{equation}
\label{64}
C(z)=|z|_p^{\alpha-1}\pi_{1}(z), \quad z\in \bQ_p^*.
\end{equation}
In this case we say that an {\it automodel\/} function $\rho(x)$ has the
degree $\pi_{\alpha}$. In particular, if
$\pi_{\alpha}(z)=|z|_p^{\alpha-1}$ we say that the {\it automodel\/}
function has the degree $\alpha-1$.

If an {\it automodel\/} function $\rho(t)$,
$t\in \bQ_p^*$ has the degree $\pi_{\alpha}$ then the {\it automodel\/}
function $|t|_p^{\beta}\rho(t)$ has the degree
$\pi_{\alpha}\pi_{0}^{-\beta}=\pi_{1}(t)|t|_p^{\alpha+\beta}$,
where $\pi_{0}(t)=|t|_p^{-1}$.

For example, the functions $|t|_p^{\alpha-1}\pi_1(t)$ and
$|t|_p^{\alpha-1}\pi_1(t)\log_{p}^{m}|t|_p$ are {\it automodel\/}
of degree $\pi_{\alpha}$.

\begin{Definition}
\label{de4} \rm
(~\cite{Kh-Sh1},~\cite{Kh-Sh2})
Let $f\in {\cD}'(\bQ_p^n)$. If there exists an {\it automodel\/} function
$\rho(t)$, $t\in \bQ_p^{*}$ of degree $\pi_{\alpha}$ such that
$$
\frac{f(tx)}{\rho(t)} \to g(x)\not\equiv 0, \quad |t|_p \to \infty,
\quad \text{in} \quad {\cD}'(\bQ_p^n).
$$
then we say that the distribution $f$ has the {\it quasi-asymptotics\/}
$g(x)$ of degree $\pi_{\alpha}$ at infinity with respect to $\rho(t)$,
and write
$$
f(x) \stackrel{{\cD}'}{\sim} g(x), \quad |x|_p \to \infty \ \big(\rho(t)\big).
$$
If for any $\alpha$ we have
$$
\frac{f(tx)}{|t|_p^{\alpha}} \to 0, \quad |t|_p \to \infty,
\quad \text{in} \quad {\cD}'(\bQ_p^n)
$$
then we say that the distribution $f$ has a {\it quasi-asymptotics\/}
of degree $-\infty$ at infinity and write $f(x) \stackrel{{\cD}'}{\sim} 0$, \
$|x|_p \to \infty$.
\end{Definition}

\begin{Lemma}
\label{lem6}
{\rm (~\cite{Kh-Sh1},~\cite{Kh-Sh2})}
Let $f\in {\cD}'(\bQ_p^n)$. If $f(x)\stackrel{{\cD}'}{\sim} g(x)\not\equiv 0$,
as $|x|_p \to \infty$ with respect to the {\it automodel\/} function
$\rho(t)$ of degree $\pi_{\alpha}$ then $g(x)$ is a homogeneous
distribution of degree $\pi_{\alpha}$ {\rm(}with respect to
Definition~{\rm\ref{de1}.(b)}{\rm)}.
\end{Lemma}

\begin{proof}
This lemma is proved by repeating the corresponding assertion
from the book~\cite{Vl-D-Zav} practically word for word.
Let $a\in \bQ_p^{*}$.
In view of Definition~\ref{de4.1} and (\ref{64}), we obtain
$$
\langle g(ax),\varphi(x)\rangle
=\lim_{|t|_p \to \infty}
\Bigl\langle \frac{f(tax)}{\rho(t)},\varphi(x)\Bigr\rangle
\qquad\qquad\qquad\qquad\qquad\qquad\qquad\quad
$$
$$
\qquad
=\pi_{\alpha}(a)\lim_{|t|_p \to \infty}
\Bigl\langle \frac{f(tax)}{\rho(ta)},\varphi(x)\Bigr\rangle
=\pi_{\alpha}(a)\langle g(x),\varphi(x)\rangle,
$$
for all $a\in \bQ_p^{*}$, \ $\varphi \in {\cD}(\bQ_p^n)$. Thus
$g(ax)=\pi_{\alpha}(a)g(x)$ for all $a\in \bQ_p^{*}$.
\end{proof}

For $n=1$, as it follows from the theorem describing all
one-dimensional {\it homogeneous} distributions
~\cite[Ch.II,\S 2.3.]{G-Gr-P},~\cite[VIII.1.]{Vl-V-Z},
and Lemma~\ref{lem6}, if $f(x)\in {\cD}'(\bQ_p)$ has the
quasi-asymptotics of degree $\pi_{\alpha}$ at infinity then
\begin{equation}
\label{65}
f(x)\stackrel{{\cD}'}{\sim} g(x)=\left\{
\begin{array}{lcr}
C\pi_{\alpha}(x), && \pi_{\alpha}\ne \pi_{0}=|x|_p^{-1}, \\
C\delta(x),       && \pi_{\alpha}=\pi_{0}=|x|_p^{-1},  \\
\end{array}
\right.
\quad |x|_p \to \infty,
\end{equation}
where $C$ is a constant, and the distribution $\pi_{\alpha}(x)$
is defined by (\ref{24}).

\begin{Definition}
\label{de5} \rm
(~\cite{Kh-Sh1},~\cite{Kh-Sh2})
Let $f\in {\cD}'(\bQ_p^n)$. If there exists an {\it automodel\/} function
$\rho(t)$, $t\in \bQ_p^{*}$ of degree $\pi_{\alpha}$ such that
$$
\frac{f\big(\frac{x}{t}\big)}{\rho(t)} \to g(x)\not\equiv 0,
\quad |t|_p \to \infty,
\quad \text{in} \quad {\cD}'(\bQ_p^n)
$$
then we say that the distribution $f$ has the {\it quasi-asymptotics\/}
$g(x)$ of degree $\big(\pi_{\alpha}\big)^{-1}$ at zero with respect to
$\rho(t)$, and write
$$
f(x) \stackrel{{\cD}'}{\sim} g(x), \quad |x|_p \to 0 \ \big(\rho(t)\big).
$$
If for any $\alpha$ we have
$$
\frac{f\big(\frac{x}{t}\big)}{|t|_p^{\alpha}} \to 0,
\quad |t|_p \to \infty,
\quad \text{in} \quad {\cD}'(\bQ_p^n)
$$
then we say that the distribution $f$ has a {\it quasi-asymptotics\/}
of degree $-\infty$ at zero, and write $f(x) \stackrel{{\cD}'}{\sim} 0$, \
$|x|_p \to 0$.
\end{Definition}

\begin{Example}
\label{ex3} \rm
Let $f_{m}\in {\cD}'(\bQ_p)$ be an {\it associated homogeneous
{\rm(}in the wide sense{\rm)}\/} distribution of degree~$\pi_{\alpha}(x)$
and order~$m$ defined by (\ref{19.3}), (\ref{19.5}).
In view of Definition~\ref{de1.1}, we have the asymptotic formulas:
$$
f_{m}(tx)=\pi_{1}(t)|t|_p^{\alpha-1}f_m(x)
\qquad\qquad\qquad\qquad\qquad\qquad\qquad\qquad\qquad
$$
$$
\qquad\qquad
+\sum_{j=1}^{m}\pi_{1}(t)|t|_p^{\alpha-1}\log_p^j|t|_pf_{m-j}(x),
\quad |t|_p \to \infty,
$$
$$
f_{m}\Big(\frac{x}{t}\Big)=\pi_{1}^{-1}(t)|t|_p^{-\alpha+1}f_m(x)
\qquad\qquad\qquad\qquad\qquad\qquad\qquad\qquad
$$
$$
\qquad
+\sum_{j=1}^{m}(-1)^{j}\pi_{1}^{-1}(t)|t|_p^{-\alpha+1}\log_p^j|t|_pf_{m-j}(x),
\quad |t|_p \to \infty.
$$
Here the coefficients of the {\it leading term\/} of both
asymptotics are homogeneous distributions $f_0$ and $(-1)^m f_0$
of degree~$\pi_{\alpha}(x)$ defined by the relation from
Definition~\ref{de1.1}.

According to the last relations and Definitions~\ref{de4},~\ref{de5},
one can easily see that
$$
\begin{array}{rclrclrcl}
\displaystyle
f_{m}(x) &\stackrel{{\cD}'}{\sim}& f_0(x), &&|x|_p \to \infty
&& \big(|t|_p^{\alpha-1}\pi_{1}(t)\log_{p}^m|t|_p\big), \smallskip \\
\displaystyle
f_{m}(x) &\stackrel{{\cD}'}{\sim}& (-1)^m f_0(x), &&|x|_p \to 0
&& \big(|t|_p^{-\alpha+1}\pi_{1}^{-1}(t)\log_{p}^m|t|_p\big).
\end{array}
$$
\end{Example}

\section{The Tauberian theorems}
\label{s7}

\begin{Theorem}
\label{th5}
{\rm (~\cite{Kh-Sh2})}
A distribution $f\in {\cD}'(\bQ_p^n)$ has a quasi-asymptotics of
degree $\pi_{\alpha}$ at infinity with respect to the automodel
function $\rho(t)$, $t\in \bQ_p^*$, i.e.,
$$
f(x) \stackrel{{\cD}'}{\sim} g(x), \quad |x|_p \to \infty
\ \big(\rho(t)\big)
$$
if and only if its Fourier transform has a quasi-asymptotics of
degree $\pi_{\alpha}^{-1}\pi_{0}^{n}=\pi_{\alpha+n}^{-1}$ at zero
with respect to the automodel function $|t|_p^n\rho(t)$, i.e.,
$$
F[f(x)](\xi) \stackrel{{\cD}'}{\sim} F[g(x)](\xi), \quad |\xi|_p \to 0
\ \big(|t|_p^n\rho(t)\big).
$$
\end{Theorem}

\begin{proof}
Let us prove the necessity. Let $f(x) \stackrel{{\cD}'}{\sim} g(x)$, \
$|x|_p \to \infty$ \ $\big(\rho(t)\big)$, i.e.,
\begin{equation}
\label{43}
\lim_{|t|_p \to \infty}
\Bigl\langle \frac{f(tx)}{\rho(t)}, \varphi(x)\Bigr\rangle
=\langle g(x), \varphi(x)\rangle, \quad \forall \  \varphi \in {\cD}(\bQ_p^n),
\end{equation}
where $\rho(t)$ is an automodel function of degree $\pi_{\alpha}$.
In view of formula (\ref{14}),
$F[f(x)](\frac{\xi}{t})=|t|_p^nF[f(tx)](\xi)$,
\ $x,\xi\in \bQ_p^n$, \ $t\in \bQ_p^*$, we have
$$
\Bigl\langle F[f(x)]\Big(\frac{\xi}{t}\Big),\varphi(\xi)\Bigr\rangle
=|t|_p^n\bigl\langle F[f(tx)](\xi),\varphi(\xi)\bigr\rangle
=|t|_p^n\bigl\langle f(tx),F[\varphi(\xi)](x)\bigr\rangle,
$$
$\varphi \in {\cD}(\bQ_p^n)$.
Hence, taking into account relation (\ref{43}), we obtain
$$
\lim_{|t|_p \to \infty}
\Bigl\langle \frac{F[f(x)](\frac{\xi}{t})}{|t|_p^n\rho(t)},
\varphi(\xi)\Bigr\rangle
=\lim_{|t|_p \to \infty}
\Bigl\langle \frac{f(tx)}{\rho(t)},F[\varphi(\xi)](x)\Bigr\rangle
\qquad
$$
$$
=\bigl\langle g(x),F[\varphi(\xi)](x)\bigr\rangle
=\bigl\langle F[g(x)](\xi),\varphi(\xi)\bigr\rangle,
\quad \forall \ \varphi \in {\cD}(\bQ_p^n),
$$
i.e., the distribution $F[f(x)](\xi)$ has the quasi-asymptotics
$F[g(x)](\xi)$ of degree $\pi_{\alpha+n}^{-1}$ at zero with respect
to $|t|_p^n\rho(t)$.

The sufficiency can be proved similarly.
\end{proof}

For $n=1$ Theorem~\ref{th5}, Lemma~\ref{lem6},
and formula (\ref{60}) imply the following corollary.

\begin{Corollary}
\label{cor6}
A distribution $f\in {\cD}'(\bQ_p)$ has a quasi-asymptotics of
degree $\pi_{\alpha}(x)$ at infinity, i.e.,
\begin{equation}
\label{43*}
f(x) \stackrel{{\cD}'}{\sim}
g(x)=\left\{
\begin{array}{lcr}
C|x|_p^{\alpha-1}\pi_1(x),
&& \pi_{\alpha}\ne \pi_{0}=|x|_p^{-1}, \\
C\delta(x),
&& \pi_{\alpha}=\pi_{0}=|x|_p^{-1},  \\
\end{array}
\right.
\quad |x|_p \to \infty,
\end{equation}
if and only if its Fourier transform $F[f]$ has a
quasi-asymptotics of degree $\pi_{\alpha+1}^{-1}(\xi)$
at zero, i.e.,
$$
F[f(x)](\xi) \stackrel{{\cD}'}{\sim}F[g(x)](\xi)
\qquad\qquad\qquad\qquad\qquad\qquad\qquad\qquad\qquad\qquad
$$
$$
\qquad\qquad
=\left\{
\begin{array}{lcr}
C\Gamma_p(\pi_{\alpha})|\xi|_p^{-\alpha}\pi_1^{-1}(\xi),
&& \pi_{\alpha}\ne \pi_{0}=|x|_p^{-1}, \\
C,
&& \pi_{\alpha}=\pi_{0}=|x|_p^{-1},  \\
\end{array}
\right.
\quad |\xi|_p \to 0,
$$
where the distribution $\pi_{\alpha}(x)=|x|_p^{\alpha-1}\pi_1(x)$
is given by {\rm (\ref{24})}.
\end{Corollary}

\begin{Theorem}
\label{th7}
Let $f \in \Phi_{\times}'(\bQ_p^n)$. Then
$$
f(x) \stackrel{\Phi_{\times}'}{\sim} g(x), \quad |x|_p \to \infty
\quad \big(\rho(t)\big)
$$
if and only if
$$
D^{\beta}_{\times}f(x) \stackrel{\Phi_{\times}'}{\sim} D^{\beta}_{\times}g(x),
\quad |x|_p \to \infty \quad \big(|t|_p^{|-\beta|}\rho(t)\big),
$$
where $\beta=(\beta_1,\dots,\beta_n)\in \bC^n$, \
$|\beta|=\beta_1+\cdots+\beta_n$.
\end{Theorem}

\begin{proof}
Let $\beta_j \ne -1$, $j=1,2,\dots$. In this case the Riesz kernel
$f_{-\beta}(x)$ is a {\it homogeneous\/} distribution of degree~$|-\beta|-n$.
According to Lemma~\ref{lem4} and formulas (\ref{58}), (\ref{59}), (\ref{62}),
we have
$$
\bigl\langle \big(D^{\beta}_{\times}f\big)(tx),\phi(x)\bigr\rangle
=\bigl\langle \big(f*f_{-\beta}\big)(tx),\phi(x)\bigr\rangle
\qquad\qquad\qquad\qquad\qquad\qquad
$$
$$
=|t|_p^{-n}
\Bigl\langle f(x),\Bigl\langle f_{-\beta}(y),\phi\Big(\frac{x+y}{t}\Big)
\Bigr\rangle\Bigr\rangle
=|t|_p^{n}
\bigl\langle f(tx),
\bigl\langle f_{-\beta}(ty),\phi(x+y)\bigr\rangle \bigr\rangle
$$
$$
=|t|_p^{|-\beta|}
\bigl\langle f(tx),
\bigl\langle f_{-\beta}(y),\phi(x+y)\bigr\rangle \bigr\rangle
=|t|_p^{|-\beta|}
\bigl\langle f(tx),\big(D^{\beta}_{\times}\phi\big)(x)\bigr\rangle,
$$
for all $\phi \in \Phi_{\times}(\bQ_p^n)$.
Thus
$$
\Bigl\langle \frac{\big(D^{\beta}_{\times}f\big)(tx)}{|t|_p^{|-\beta|}\rho(t)},
\phi(x)\Bigr\rangle
=\Bigl\langle \frac{f(tx)}{\rho(t)},
\big(D^{\beta}_{\times}\phi\big)(x)\Bigr\rangle.
$$

Next, passing to the limit in the above relation, as
$|t|_p \to \infty$, we obtain
$$
\lim_{|t|_p \to \infty}
\Bigl\langle \frac{\big(D^{\beta}_{\times}f\big)(tx)}{|t|_p^{|-\beta|}\rho(t)},
\phi(x)\Bigr\rangle
=\lim_{|t|_p \to \infty}
\Bigl\langle \frac{f(tx)}{\rho(t)},
\big(D^{\beta}_{\times}\phi\big)(x)\Bigr\rangle
$$
That is, $\lim_{|t|_p \to \infty}
\frac{(D^{\beta}_{\times}f)(tx)}{|t|_p^{|-\beta|}\rho(t)}
=D^{\beta}_{\times}g(x)$
in $\Phi_{\times}'(\bQ_p^n)$ if and only if
$\lim_{|t|_p \to \infty}\frac{f(tx)}{\rho(t)}=g(x)$ in
$\Phi_{\times}'(\bQ_p^n)$.
Thus this case of the theorem is proved.

Consider the case where among all $\beta_1,\dots,\beta_n$ there are $k$
pieces such that $=-1$ and $n-k$ pieces such that $\ne -1$. In this case
the Riesz kernel $f_{-\beta}(x)$ is an {\it associated homogeneous\/}
distribution of degree~$|-\beta|-n$ and order $k$, \ $k=1,\dots,n$.
Let $\beta_1=\cdots=\beta_k=-1$, \  $\beta_{k+1},\cdots,\beta_{n}\ne -1$.
Then according to (\ref{58.1}),
$$
f_{-\beta}(ty)=|t|_p^{|-\beta|-n}
(-1)^k\frac{(p-1)^k}{\log^k p}(\log|y_1|_p+\log|t|_p)\times
\qquad\qquad\qquad
$$
$$
\cdots\times(\log|y_k|_p+\log|t|_p)
\times\frac{|y_{k+1}|_p^{-\beta_{k+1}-1}}{\Gamma_p(-\beta_{k+1})}
\times\cdots\times \frac{|y_n|_p^{-\beta_{n}-1}}{\Gamma_p(-\beta_n)}
$$
$$
=|t|_p^{|-\beta|-n}f_{-\beta}(y)
\qquad\qquad\qquad\qquad\qquad\qquad\qquad\quad
$$
$$
\qquad\qquad\quad
+|t|_p^{|-\beta|-n}(-1)^k\frac{(p-1)^k}{\log^k p}
\frac{|y_{k+1}|_p^{-\beta_{k+1}-1}}{\Gamma_p(-\beta_{k+1})}
\times\cdots\times\frac{|y_n|_p^{-\beta_{n}-1}}{\Gamma_p(-\beta_n)}
$$
$$
\times
\bigg(\Big(\log|y_2|_p\times\cdots\times\log|y_k|_p
+\cdots+\log|y_1|_p\times\cdots\times\log|y_{k-1}|_p\Big)\log|t|_p
$$
\begin{equation}
\label{43**}
+\cdots+\Big(\log|y_1|_p+\cdots+\log|y_k|_p\Big)\log^{k-1}|t|_p
+\log^{k}|t|_p\bigg).
\end{equation}

It is easy to verify that in view of characterization (\ref{50}),
$$
\bigl\langle f_{-\beta}(ty), \phi(x+y)\bigr\rangle
=|t|_p^{|-\beta|-n}\bigl\langle f_{-\beta}(y), \phi(x+y)\bigr\rangle
\qquad\qquad\qquad
$$
\begin{equation}
\label{43***}
\qquad\qquad\qquad
=|t|_p^{|-\beta|-n}\big(D^{\beta}_{\times}\phi\big)(x),
\quad \phi \in \Phi_{\times}(\bQ_p^n).
\end{equation}

For example, taking into account (\ref{50}), we obtain
$$
\Bigl\langle
\times_{j=2}^{k}\log|x_j-y_j|_p\times
\times_{i=k+1}^{n}|x_{i}-y_{i}|_p^{-\beta_{i}-1},
\int_{\bQ_p}\phi(y_1,y_2,\dots,y_n)\,dy_1 \Bigr\rangle=0,
$$
for all $\phi \in \Phi_{\times}(\bQ_p^n)$.
In a similar way, one can prove that all terms in (\ref{43**}),
with the exception of $|t|_p^{|-\beta|-n}f_{-\beta}(y)$,
{\it do not give any contribution\/} to the functional
$\langle f_{-\beta}(ty),\phi(x+y)\rangle$, where
$\beta_1=\cdots=\beta_k=-1$, \  $\beta_{k+1},\cdots,\beta_{n}\ne -1$.
Thus repeating the above calculations almost word for word and using
(\ref{43***}), we prove this case of the theorem.
\end{proof}

\begin{Theorem}
\label{th8}
Let $f \in \Phi'(\bQ_p^n)$. Then
$$
f(x) \stackrel{\Phi'}{\sim} g(x), \quad |x|_p \to \infty
\quad \big(\rho(t)\big)
$$
if and only if
$$
D^{\beta}f(x) \stackrel{\Phi'}{\sim} D^{\beta}g(x),
\quad |x|_p \to \infty \quad \big(|t|_p^{-\beta}\rho(t)\big),
$$
where $\beta\in \bC$.
\end{Theorem}

\begin{proof}
Let $\beta \ne -n$, $j=1,2,\dots$. Since the Riesz kernel
$\kappa_{-\beta}(x)$ is a {\it homogeneous\/} distribution of
degree~$-\beta-n$, according to Lemma~\ref{lem4.1} and formulas
(\ref{63.4}), (\ref{59**}), (\ref{62**}),
we have
$$
\bigl\langle \big(D^{\beta}f\big)(tx),\phi(x)\bigr\rangle
=\bigl\langle (f*\kappa_{-\beta}\big)(tx),\phi(x)\bigr\rangle
\qquad\qquad\qquad\qquad\qquad\qquad\quad
$$
$$
=|t|_p^{-n}
\Bigl\langle f(x),\Bigl\langle \kappa_{-\beta}(y),\phi\Big(\frac{x+y}{t}\Big)
\Bigr\rangle\Bigr\rangle
=|t|_p^{n}
\bigl\langle f(tx),
\bigl\langle \kappa_{-\beta}(ty),\phi(x+y)\bigr\rangle \bigr\rangle
$$
$$
=|t|_p^{-\beta}
\bigl\langle f(tx),
\bigl\langle \kappa_{-\beta}(y),\phi(x+y)\bigr\rangle \bigr\rangle,
=|t|_p^{-\beta}
\bigl\langle f(tx),\big(D^{\beta}\phi\big)(x)\bigr\rangle,
\qquad\qquad
$$
for all $\phi \in \Phi(\bQ_p^n)$.

Passing to the limit in the above relation, as
$|t|_p \to \infty$, we obtain
$$
\lim_{|t|_p \to \infty}
\Bigl\langle \frac{\big(D^{\beta}f\big)(tx)}{|t|_p^{-\beta}\rho(t)},
\phi(x)\Bigr\rangle
=\lim_{|t|_p \to \infty}
\Bigl\langle \frac{f(tx)}{\rho(t)},\big(D^{\beta}\phi\big)(x)\Bigr\rangle
$$
Thus this case of the theorem is proved.

Let $\beta=-n$. In this case the Riesz kernel $\kappa_{n}(x)$ is an
{\it associated homogeneous\/} distribution of degree~$0$ and order $1$.
According to (\ref{63.7}), we have
$$
\kappa_{n}(ty)=-\frac{1-p^{-n}}{\log p}\log|y|_p
-\frac{1-p^{-n}}{\log p}\log|t|_p.
$$
In view of (\ref{54}),
$$
\bigl\langle \kappa_{n}(ty), \phi(x+y)\bigr\rangle
=\bigl\langle \kappa_{n}(y), \phi(x+y)\bigr\rangle
\qquad\qquad\qquad\qquad\qquad\qquad\qquad
$$
$$
-\frac{1-p^{-n}}{\log p}\log|t|_p
\bigl\langle 1, \phi(x+y)\bigr\rangle
=\big(D^{-n}\phi\big)(x),
\quad \phi \in \Phi(\bQ_p^n).
$$
Thus repeating the above calculations almost word for word and using
the last relation, we prove this case of the theorem.
\end{proof}

\begin{Theorem}
\label{th9}
A distribution $f\in \Phi'(\bQ_p)$ has a quasi-asymptotics at infinity
with respect to an automodel function $\rho(t)$ of degree $\pi_{\alpha}$
if and only if there exists a positive integer $N>-\alpha+1$ such that
$$
\lim_{|x|_p \to \infty}\frac{D^{-N}f(x)}{|x|_p^{N}\rho(x)}=A \ne 0,
$$
i.e., the {\rm(}fractional{\rm)} primitive $D^{-N}f(x)$ of
order $N$ has an asymptotics at infinity {\rm(}understood in the usual
sense{\rm)} of degree $\pi_{\alpha+N}$.
\end{Theorem}

\begin{proof}
By setting $\beta=-N$, $N>-\alpha+1$ in Theorem~\ref{th7}, we obtain that
relation (\ref{43*}) holds if and only if
\begin{equation}
\label{80}
D^{-N}f(x)\stackrel{\Phi'}{\sim}D^{-N}g(x)
=C\left\{
\begin{array}{lcr}
D^{-N}\big(|x|_p^{\alpha-1}\pi_1(x)\big), && \pi_{\alpha}\ne \pi_{0}, \\
D^{-N}\big(\delta(x)\big), && \pi_{\alpha}=\pi_{0},  \\
\end{array}
\right.
\end{equation}
as $|x|_p \to \infty$ \, $\big(|t|_p^{N}\rho(t)\big)$,
where $\pi_{0}=|x|_p^{-1}$.

If $\pi_{\alpha}\ne \pi_{0}=|x|_p^{-1}$, with the help of formulas
(\ref{25.4}), (\ref{25.5}), (\ref{57.2}), we find that
\begin{equation}
\label{81}
D^{-N}g(x)=C\frac{|x|_p^{N-1}}{\Gamma_p(N)}*\big(|x|_p^{\alpha-1}\pi_{1}(x)\big)
=C\frac{\Gamma_p(\pi_{\alpha})}{\Gamma_p(\pi_{\alpha+N})}
|x|_p^{\alpha+N-1}\pi_{1}(x),
\end{equation}
where the $\Gamma$-functions are given by (\ref{25.1}), (\ref{25}).
If $\pi_{\alpha}=\pi_{0}=|x|_p^{-1}$ then
\begin{equation}
\label{82}
D^{-N}g(x)=C\frac{|x|_p^{N-1}}{\Gamma_p(N)}*\delta(x)
=C\frac{|x|_p^{N-1}}{\Gamma_p(N)}.
\end{equation}

Formulas (\ref{80}), (\ref{81}), (\ref{82}) imply that
$$
\lim_{|t|_p \to \infty}
\Bigl\langle \frac{\big(D^{-N}f\big)(tx)}{|t|_p^{N}\rho(t)},
\phi(x)\Bigr\rangle
=C\frac{\Gamma_p(\pi_{\alpha})}{\Gamma_p(\pi_{\alpha+N})}
\bigl\langle |x|_p^{\alpha+N-1}\pi_{1}(x),\phi(x)\bigr\rangle,
$$
for all $\phi \in \Phi(\bQ_p)$. Since $\alpha+N-1>0$, we have
\begin{equation}
\label{83}
\lim_{|t|_p \to \infty}
\frac{\big(D^{-N}f\big)(tx)}{|t|_p^{N}\rho(t)}
=C\frac{\Gamma_p(\pi_{\alpha})}{\Gamma_p(\pi_{\alpha+N})}
|x|_p^{\alpha+N-1}\pi_{1}(x).
\end{equation}

By using Definition~\ref{de4.1} and formula (\ref{64}),
relation (\ref{83}) can be rewritten in the following form
$$
A=C\frac{\Gamma_p(\pi_{\alpha})}{\Gamma_p(\pi_{\alpha+N})}
=\lim_{|t|_p \to \infty}
\frac{\big(D^{-N}f\big)(tx)}{|t|_p^{N}\rho(t)|x|_p^{\alpha+N-1}\pi_{1}(x)}
\qquad\qquad\qquad\qquad\qquad
$$
$$
=\lim_{|tx|_p \to \infty}
\frac{\big(D^{-N}f\big)(tx)}{|tx|_p^{N}\rho(tx)}
\lim_{|t|_p \to \infty}
\frac{\rho(tx)}{|x|_p^{\alpha-1}\pi_{1}(x)\rho(t)}
=\lim_{|y|_p \to \infty}
\frac{\big(D^{-N}f\big)(y)}{|y|_p^{N}\rho(y)}.
$$
\end{proof}

\begin{Theorem}
\label{th10}
Let $\cA(\xi)\in \cE(\bQ_p^n\setminus \{0\})$ be the symbol of
a {\it homogeneous\/} pseudo-differential operator $A$ of degree~$\pi_{\beta}$,
and $f \in \Phi'(\bQ_p^n)$. Then
$$
f(x) \stackrel{\Phi'}{\sim} g(x), \quad |x|_p \to \infty
\quad \big(\rho(t)\big)
$$
if and only if
$$
(Af)(x) \stackrel{\Phi'}{\sim} (Ag)(x),
\quad |x|_p \to \infty \quad \big(\pi_{\beta}^{-1}(t)\rho(t)\big).
$$
\end{Theorem}

\begin{proof}
Since the Lizorkin space $\Phi(\bQ_p^n)$ is invariant under the
pseudo-differential operator $A$ (see Sec.~\ref{s5}),
according to formulas (\ref{64.3*}), (\ref{64.5}), and
(\ref{14}), (\ref{17}), we have
$$
\bigl\langle \big(Af\big)(tx),\phi(x)\bigr\rangle
=|t|_p^{-n}\Bigl\langle f(x),A^{T}\phi\Big(\frac{x}{t}\Big)\Bigr\rangle
\qquad\qquad\qquad\qquad\qquad
$$
$$
=|t|_p^{-n}\Bigl\langle f(x),
F^{-1}\big[\cA(-\xi)F[\phi\Big(\frac{x}{t}\Big)](\xi)\big](x)\Bigr\rangle
\qquad\qquad\quad
$$
$$
=\frac{1}{\pi_{\beta}(t)}\bigl\langle f(x),
F^{-1}\big[\cA(-t\xi)F[\phi(x)](t\xi)\big](x)\bigr\rangle
\qquad\qquad\quad
$$
$$
=\frac{|t|_p^{-n}}{\pi_{\beta}(t)}\Bigl\langle f(x),
F^{-1}\big[\cA(-\xi)F[\phi(x)](\xi)\big]\Big(\frac{x}{t}\Big)\Bigr\rangle
\qquad\qquad\quad
$$
$$
\qquad
=\frac{1}{\pi_{\beta}(t)}\bigl\langle f(tx),
F^{-1}\big[\cA(-\xi)F[\phi(x)](\xi)\big](x)\bigr\rangle,
\quad \forall \, \phi \in \Phi(\bQ_p^n).
$$

Passing to the limit in the above relation, as
$|t|_p \to \infty$, we obtain
$$
\lim_{|t|_p \to \infty}
\Bigl\langle \frac{(Af)(tx)}{\pi_{\beta}^{-1}(t)\rho(t)},
\phi(x)\Bigr\rangle
=\lim_{|t|_p \to \infty}
\Bigl\langle \frac{f(tx)}{\rho(t)},\big(A^{T}\phi\big)(x)\Bigr\rangle,
$$
i.e., in view of (\ref{64.5}),
$\lim_{|t|_p \to \infty}\frac{(Af)(tx)}{\pi_{\beta}^{-1}(t)\rho(t)}=Ag(x)$
in $\Phi'(\bQ_p^n)$ if and only if
$\lim_{|t|_p \to \infty}\frac{f(tx)}{\rho(t)}=g(x)$ in
$\Phi'(\bQ_p^n)$.
Thus the theorem is proved.
\end{proof}

\begin{center}
{\bf Acknowledgements }
\end{center}
The authors would like to thank Yu.~N.~Drozzinov, S.~V.~Kozyrev,
I.~V.~Volovich, B.~I.~Zavialov for fruitful discussions.

\end{document}